\documentclass[aps,pra,reprint,showpacs,preprintnumbers,amsmath,amssymb,floatfix]{revtex4-1}

\input{epsf}

\font\bm=cmmib10 at 10pt
\font\bms=cmmib10 at 7pt \textfont9=\bm \scriptfont9=\bms
\mathchardef\balpha= "790B
\mathchardef\bbeta= "790C
\mathchardef\bTheta= "7902
\mathchardef\bzeta= "7910
\mathchardef\bOmega= "790A
\mathchardef\bGamma= "7900
\mathchardef\bDelta= "7901
\mathchardef\bPhi= "7908
\mathchardef\bphi= "791E
\mathchardef\bomega= "7921
\mathchardef\bxi= "7918
\mathchardef\bet= "7911
\mathchardef\brho= "791A
\mathchardef\btau= "791C
\mathchardef\bmu= "7916
\mathchardef\bvarpi= "7924
\def \Tr {\hbox {Tr}}

\def \cmt {{\copyright}}
\def \lvec{(\kern-.26em(}
\def\pmb#1{\setbox0=\hbox{#1}%
\def \lvec{(\kern-.26em(}
\kern-.025em\copy0\kern-\wd0
\kern.05em\copy0\kern-\wd0
\kern-.025em\raise.0433em\box0 }
\mathchardef\btheta= "7912
\usepackage{amsmath,graphicx}
\begin{document}
\title{Spin-Velocity Correlations of Optically Pumped Atoms}
\author{R. Marsland III}
\altaffiliation[Present address: ]{Condensed Matter Theory Group, Department of Physics, Massachusetts Institute of Technology, Cambridge, Massachusetts 02139, USA}
\author{B. H. McGuyer}
\altaffiliation[Present address: ]{Department of Physics, Columbia University, New York, New York 10027, USA}
\author{B. A. Olsen}
\altaffiliation[Present address: ]{Department of Physics and Astronomy and Rice Quantum Institute, Rice University, Houston, Texas 77005, USA}
\author{W. Happer}
\affiliation{%
Joseph Henry Laboratory, Department of Physics, Princeton University, Princeton, New Jersey 08544, USA}
\date{\today}

\begin{abstract}
We present efficient theoretical tools for describing the optical pumping of atoms by light propagating at arbitrary directions with respect to an external magnetic field, at buffer-gas pressures that are small enough for velocity-selective optical pumping (VSOP) but large enough to cause substantial collisional relaxation of the velocities and the spin. These are the conditions for the sodium atoms at an altitude of about 100~km that are used as guidestars for adaptive optics in modern ground-based telescopes to correct for aberrations due to atmospheric turbulence. We use spin and velocity relaxation modes to describe the distribution of atoms in spin space (including both populations and coherences) and velocity space. Cusp kernels are used to describe velocity-changing collisions. Optical pumping operators are represented as a sum of poles in the complex velocity plane. Signals simulated with these methods are in excellent agreement with previous experiments and with preliminary experiments in our laboratory.
\end{abstract}

\maketitle
\section{Introduction}
In this paper we discuss efficient ways to model optically pumped atoms in a regime where velocity-selective optical pumping (VSOP) is possible, but where collisional rates with buffer gases are too high to permit the use of models for cooling and trapping in the near absence of collisions. This is the regime of sodium guidestar atoms. These naturally-occurring layers of sodium atoms at altitudes of 90--100~km above the Earth's surface \cite{Happer94} are illuminated by ground-based lasers, and the returning photons
are used to measure the relative retardation of wave fronts across an optical aperture. This retardation information can be used with a deformable mirror to correct for the
aberrations from atmospheric turbulence and to allow the receiving optics to produce a more nearly diffraction-limited image of astronomical objects.

The performance of guidestar systems is limited by the loss of atoms from the most strongly backscattering spin sublevels and velocity groups. The most important reasons for these losses are: collisions with the residual atmospheric gases, which transfer atoms from strongly absorbing to weakly absorbing spin sublevels, or which shift the atoms into velocity groups that are not in resonance with the pumping light; Larmor precession of the spins away from strongly absorbing orientations if the geomagnetic field is not parallel to the direction of the laser beam; and unwanted optical pumping into weakly absorbing sublevels. The powerful modeling methods discussed here make it easier to explore the parameter space of these processes and to optimize the performance of guidestar systems. These methods also provide a more realistic and numerically convenient way to model laboratory experiments with VSOP of atoms in low-pressure buffer gases. In contrast to previous work on this topic, for example references \cite{Aminoff82, Quivers86, Tomasi93}, we account for the full hyperfine structure of real alkali-metal atoms, we show how to use spin-relaxation modes \cite{Bouchiat63a, Bouchiat63b} to incorporate into the model the complicated spin relaxation of Na guidestar atoms due to collisions with paramagnetic oxygen atoms, and we use recently developed cusp kernels \cite{McGuyer12} to realistically and efficiently model velocity-changing collisions.

This article is organized as follows.
In Section \ref{DenMat}, we introduce the Liouville space of the coupled spin and velocity distributions of the atoms.
In Section \ref{spin} we introduce spin-relaxation modes $s_j$~\cite{Bouchiat63a,Bouchiat63b} to describe the spin distributions, and we introduce the concept of conjugate spin-mode indices, $j$ and $j^*$. The amplitudes $|\chi_j)$ of the spin modes in velocity space provide a complete description of the spin and velocity polarization of the atoms. The little-known Liouville conjugate operation, denoted by the superscript $^{\ddag}$, and the transposition operator $T$ for Liouville-space operators are discussed in Section \ref{spinB}. Using Liouville conjugates reduces the numerical computing requirements by nearly a factor of 2.

In Section \ref{velocity} we show how to describe velocity distributions with velocity-relaxation modes $|v_n)$ \cite{Snider86, Morgan10}. We use the velocity modes to show that a simple transformation of the widely used Keilson-Storer kernels \cite{Keilson52} leads to much more realistic and useful cusp kernels \cite{McGuyer12} for describing velocity-changing collisions.
In Section \ref{pressure} we present a simple model for the transition from collision-free, ballistic flight to any container walls at very low buffer-gas pressure to diffusional wall losses at higher pressure. We sketch how the relative sizes of the laser beam and the cell affect these processes. In an Appendix we show how to deduce the rate of velocity-changing collisions, $\gamma_{\rm vd}$ from the spatial diffusion coefficient $D$ and the smallest-nonzero eigenvalue $\alpha_1$ of the collision operator $A_{\rm vd}$ with the little-known formula (\ref{sd80}). In (\ref{rid16}) of Section \ref{dark}, we show that spin-changing and velocity-changing collisions cause the spin-mode amplitudes $|\chi_j)$ to relax exponentially in time at the rate $K_j$, where $K_j$ is a kernel in velocity space.

In Section \ref{pumping}, we introduce a velocity-dependent optical pumping operator $A_{\rm op}$, which we write as the sum of poles in the complex-velocity plane at locations determined by the laser frequency and the optical Bohr frequencies. The pole expansion facilitates velocity averages in terms of Faddeeva functions (Voigt profiles) \cite{Weideman94}. The poles have ``residue matrices'' that are independent of the laser frequency and atomic velocity. In Section VIII, we show that the steady-state mode amplitudes $|\chi_j)$ generated by the combined effects of optical pumping, spin relaxation, and velocity relaxation can be written in terms of Green's functions, $G_j\propto K_j^{-1}$. We show that if the kernel $K_j$ is a cusp kernel or linear combination thereof, $K_j^{-1}$ is also cusp kernel or linear combination thereof. Being able to invert cusp kernels in closed form greatly simplifies the numerical evaluation of the mode amplitudes $|\chi_j)$. This simplification is not possible with Keilson-Storer kernels or any other collision kernel that we know of.
We present an explicit, series solution (\ref{ss2}) for the mode amplitudes $|\chi_j)$ in powers of the pumping light intensity, with particular emphasis on the first-order solution (\ref{sr10}).

Finally, in Section IX, we use (\ref{sr10}) to demonstrate how the methods we present are in excellent agreement with existing VSOP experiments. We also describe a new type of magnetic-depolarization experiment that can be carried out under laboratory conditions and readily interpreted with the powerful modeling methods described in this paper. Such experiments would provide much more detailed experimental information about the nature of velocity-changing collisions.

\section{The Density Matrix\label{DenMat}}
For optical pumping at low buffer-gas pressure we need to account for both the spin-polarization and the velocity $v$ of atoms along the
light beam. We introduce a dimensionless velocity,
\begin{equation}
x=\frac{v}{v_D},
\label{op2}
\end{equation}
where the most probable speed $v_D$ along the laser beam
is given by
\begin{equation}
v_D=\sqrt{\frac{2k_BT}{M}}.
\label{op4}
\end{equation}
Here $T$ is the absolute temperature, $k_B$ is Boltzmann's constant, and $M$ is the mass of the
atom.

We write the incremental density matrix $d\rho$ for the spin-polarized, ground-state atoms
with velocities between $x$ and $x+dx$ as
\begin{equation}
d\rho=\chi(x)\, dx.
\label{op6}
\end{equation}
Here $\chi(x)$ denotes a square $g^{\{g\}}\times g^{\{g\}}$ matrix in Schr\"odinger spin space for ground-state atoms, where the dimension of the spin space is
$g^{\{g\}}=(2S+1)(2I+1)$. The electronic-spin quantum number of the ground-state atom is $S$, and the nuclear-spin quantum number is $I$.
The total probability for the atom to have some velocity and be in some spin state must be unity, so we must have
\begin{equation}
\Tr\int \chi(x)\,dx=1.
\label{op12}
\end{equation}
\subsection{Energies and energy basis states}
It will be convenient to describe the atoms in terms of the energy eigenstates $|\mu\rangle$. These are defined
by the time-independent Schr\"odinger equation
\begin{equation}
H^{\{g\}}|\mu\rangle =E_{\mu}|\mu\rangle,
\label{eeb2}
\end{equation}
which determines the energy shifts $E_{\mu}$ of the basis states $|\mu\rangle$ from their center of gravity due to hyperfine interactions and externally applied magnetic fields.
The spin Hamiltonian $H^{\{g\}}$ for the $^2$S$_{1/2}$ ground state of an alkali-metal atom is traceless and includes hyperfine couplings of the nuclear and electronic spins to each other as well as their couplings to an externally applied magnetic field. The energy sublevels $|\bar \mu\rangle$ and energy shifts $E_{\bar \mu}$ of optically-excited atoms are given in like manner by
\begin{equation}
H^{\{e\}}|\bar\mu\rangle =E_{\bar \mu}|\bar \mu\rangle.
\label{eeb10}
\end{equation}
The Bohr frequency for an optical transition from the sublevel $|\mu\rangle$ to the sublevel $|\bar\mu\rangle$ is
\begin{equation}
\omega_{\bar \mu \mu}=\omega^{\{eg\}}+\frac{E_{\bar \mu}-E_{\mu}}{\hbar}.
\label{eeb12}
\end{equation}
Here $\omega^{\{eg\}}=ck^{\{eg\}}=2\pi c/\lambda^{\{eg\}}$ is the mean value of the frequencies (\ref{eeb12}), averaged over all possible combinations of the sublevel labels, $\bar\mu$ and
$\mu$. The corresponding spatial frequency is $k^{\{eg\}}$ and wavelength is $\lambda^{\{eg\}}$.

For the low geomagnetic fields of interest to us, it will sometimes be convenient to use low-field labels of the energy sublevels
$\mu\leftrightarrow f_{\mu} m_{\mu}$. Here, $f$ denotes the approximate total spin angular momentum quantum number of the sublevel and $m$ is the exact azimuthal quantum number
along a quantization axis defined by the external magnetic field.

\subsection{Liouville space}
We use a generalization of the Liouville-space formalism of the recent book {\it Optically Pumped Atoms} by Happer, Jau and Walker \cite{OPA} (which we will refer to as OPA) for handling the large amount of information needed to describe the spin-velocity correlations of optically pumped atoms. One of the best early descriptions of Liouville space is given in the book by Ernst {\it et al.} \cite{Ernst}.

To describe the (dimensionless) velocity $x$, we can use $n_x$ evenly spaced sample velocities,
\begin{equation}
x_1, x_2, \ldots, x_{n_x},\quad\hbox{with}\quad x_{k+1}-x_k=\delta x.
\label{ls2}
\end{equation}
Then the velocity-dependent spin polarization can be defined by the elements of the spin density matrix $\chi$ of (\ref{op6}), which we will call ``spin-velocity correlations,"
\begin{equation}
\chi_{\mu\nu}(x_k)=\frac{1}{\delta x}(x_k|\chi_{\mu\nu}).
\label{ls4}
\end{equation}
We will think of $(x_k|\chi_{\mu\nu})$ as the projection onto the velocity-space basis vector $|x_k)$ of the abstract, velocity-space column vector $|\chi_{\mu\nu})$. It will be convenient to represent the total density matrix for spin-velocity space as the abstract, Kronecker-product column vector
\begin{equation}
|\Phi) =\sum_{\mu\nu}|\mu\nu)\otimes|\chi_{\mu\nu}),
\label{ls22}
\end{equation}
where $|\mu\nu)$ is the Liouville-space representation of the spin density matrix basis element $|\mu\rangle\langle\nu|$.
We turn now to the time evolution of $|\Phi)$.
\section{Spin-Damping\label{spin}}
There is negligible correlation between spin relaxation and velocity relaxation for laboratory VSOP experiments, where all the spin relaxation is due to collisions of polarized atoms with
the cell walls, or for sodium guidestar experiments where the spin relaxation is is almost all due to binary spin-exchange collisions with oxygen molecules, and where the electron spin, but not the nuclear spin of the Na atom may flip. We therefore take the spin-relaxation processes to be independent of velocity-relaxation processes and we write the rate of change of (\ref{ls22}) due to the hyperfine interaction, the externally applied magnetic field, and spin-changing collisions as
\begin{equation}
\frac{\partial}{\partial t}|\Phi)= -\sum_{\mu\nu}\Gamma|\mu\nu)\otimes |\chi_{\mu\nu}).
\label{sev2}
\end{equation}
Here the spin-damping operator is independent of velocity and is given by
\begin{equation}
\Gamma=\frac{i}{\hbar}H^{\cmt}+\gamma_\text{sd}A_\text{sd}.
\label{sev4}
\end{equation}
The evolution due to internal hyperfine couplings of the electron and nuclear spins as well as their interactions with externally applied magnetic fields is
given by the Liouville-space Hamiltonian, a ``commutator superoperator" as described by (4.85) of OPA \cite{OPA},
\begin{equation}
H^{\cmt}=\hbar\sum_{\mu\nu}\omega_{\mu\nu}|\mu\nu)(\mu\nu|,
\label{sev6}
\end{equation}
where the Bohr frequencies of the ground-state atoms are
\begin{equation}
\omega_{\mu\nu}=\frac{E_{\mu}-E_{\nu}}{\hbar}.
\label{sev8}
\end{equation}
In (\ref{sev4}) spin-changing collisions occur at a rate $\gamma_{\rm sd}$ and the details of the spin relaxation are described by the dimensionless matrix operator
\begin{equation}
A_{\rm sd}=\sum_{\mu\nu\mu'\nu'} |\mu\nu)(\mu\nu|A_{\rm sd}|\mu'\nu')(\mu'\nu'|.
\label{sev9}
\end{equation}

The specific form of the damping operator $A_{\rm sd}$ for various collisional interactions is discussed in Chapter 10 of OPA \cite{OPA}.
Regardless of the particular details, for the evolution described by (\ref{sev2}) to conserve the number of atoms, the spin-evolution operators must satisfy the constraints 
\begin{equation}
\lvec s_0|\Gamma=\lvec s_0|H^{\cmt}=\lvec s_0|A_{\rm sd}=0.
\label{sev10}
\end{equation}
The equilibrium left eigenvector, about which we will have more to say below, is
\begin{equation}
\lvec s_0|=\sum_{\mu}(\mu\mu|.
\label{sev12}
\end{equation}
\subsection{Spin Modes\label{spinA}}
In this section we discuss how to handle the complicated spin relaxation of sodium guidestar atoms due to gas-phase collisions with O$_2$ molecules and O atoms. For the weakly-relaxing buffer gases used in laboratory experiments, spin-relaxation collisions in the gas are slow enough to be neglected, and the spin relaxation is almost entirely due to collisions with the walls.
An especially convenient basis for the spin polarization under conditions of strong collisional relaxation in the gas is provided by the spin-relaxation modes. To our knowledge, spin modes were first introduced by M. A. Bouchiat \cite{Bouchiat63a, Bouchiat63b} to discuss the curious multi-exponential decays observed in velocity-independent optical pumping. The spin modes
are the right eigenvectors of the spin-damping operator (\ref{sev4}),
\begin{equation}
\Gamma|s_j)=\gamma_j|s_j).
\label{rev2}
\end{equation}
Here the symbol $|s_j)$ denotes a $g^{\{g\}2} \times 1$ column vector in Liouville space, formed from a $g^{\{g\}}\times g^{\{g\}}$ matrix $s_j$
of Schr\"odinger space, as described by (1.2) and (1.3) of OPA \cite{OPA}, by placing each column of $s_j$ below the one on its left. We will enumerate the
modes such that
\begin{equation}
|\gamma_0|=0\le |\gamma_1|\le|\gamma_2|\le \ldots \le |\gamma_{n_s-1}|.
\label{rev4}
\end{equation}
For degenerate spin modes $|s_j) \ne |s_{j'})\ne|s_{j''})...$ with $\gamma_j=\gamma_{j'}=\gamma_{j''}...$ one is free to take linear combinations
of spin modes with other distinguishing properties, often the angular momentum. A simple example of longitudinal spin modes is shown in Fig. 10.5 of OPA \cite{OPA}.

In the absence of any pumping mechanisms, the spin density matrix of the atoms will relax to the thermal equilibrium state
\begin{equation}
s_0=\frac{1}{Z}e^{-H^{\{g\}}/k_BT}\quad\hbox{or}\quad |s_0)=\frac{1}{Z}|e^{-H^{\{g\}}/k_BT}).
\label{lev8}
\end{equation}
The partition function $Z$ is
\begin{equation}
Z=\Tr\left[e^{-H^{\{g\}}/k_BT}\right].
\label{lev10}
\end{equation}
The evolution operator $\Gamma$ must cause no changes to the steady state, or
\begin{equation}
\Gamma |s_0)=0.
\label{lev12}
\end{equation}

The left eigenvectors of $\Gamma$ satisfy an eigenvalue equation analogous to (\ref{rev2}) with the same set of eigenvalues,
\begin{equation}
\lvec s_j|\Gamma=\lvec s_j|\gamma_j.
\label{lev2}
\end{equation}
We will use the symbol $\lvec s_j|$ to distinguish a left eigenvector, the solution of $(\ref{lev2})$, from the Hermitian
conjugate $(s_j|=|s_j)^{\dag}$ of the right eigenvector $|s_j)$, the solution of (\ref{rev2}) with the same eigenvalue $\gamma_j$. It is necessary to make this distinction because
there may be right and left eigenvectors for which $\lvec s_j|$ is not a simple multiple of $(s_j|=|s_j)^{\dag}$. The right eigenvectors $|s_j)$ are analogous to the three primitive vectors of a crystal. For crystals of low symmetry, which correspond to non-normal operators, the primitive vectors can be non-orthogonal. The left eigenvectors $\lvec s_j|$ are analogous to the reciprocal primitive vectors.
Except for rare, singular combinations of the parameters of (\ref{sev4}), the right eigenvectors $|s_j)$ span the spin space of the atom.  As long as the right eigenvectors are linearly independent we can choose left eigenvectors such that
\begin{equation}
\lvec s_j|s_k)=\delta_{jk}\quad\hbox{and}\quad \sum_{j}|s_j)\lvec s_j|=1^{\{s\}}.
\label{lev6}
\end{equation}
Using the left and right eigenvectors we can write the net spin-damping operator as
\begin{equation}
\Gamma =\sum_{j}\gamma_j| s_j)\lvec s_j|.
\label{lev18}
\end{equation}

\emph{High-temperature limit.}\quad
For our applications, the thermal energy $k_BT$ will normally be so large compared to the energy differences of ground-state sublevels, that it will be an excellent approximation to simplify the equilibrium spin mode (\ref{lev8}) to
\begin{equation}
s_0=\frac{1^{\{g\}}}{g^{\{g\}}}\quad\hbox{or}\quad |s_0)=\frac{|1^{\{g\}})}{g^{\{g\}}},
\label{lev20}
\end{equation}
where the unit operator for the Schr\"odinger ground state is $1^{\{g\}}=\sum|\mu\rangle\langle\mu|$.
\subsection{Hermitian conjugates of spin modes \label{spinB}}
According to (4.4.2) of OPA \cite{OPA}, $\Gamma$ must be Liouvillian,
\begin{equation}
\Gamma^{\ddag}=\Gamma,
\label{hcm2}
\end{equation}
to keep the density matrix Hermitian as it evolves in time. From this requirement, we can derive some identities that dramatically reduce the amount of computation necessary to simulate a guidestar or VSOP scenario, and that simplify the form of the evolution equations. The Liouville conjugate $\Gamma^{\ddag}$ of $\Gamma$ is defined by
\begin{equation}
\Gamma^{\ddag}=T\Gamma^*T.
\label{hcm4}
\end{equation}
The transposition operator of (\ref{hcm4}) can be written as
\begin{equation}
T=\sum_{\mu\nu}|\nu\mu)(\mu\nu|.
\label{hcm6}
\end{equation}
Substituting (\ref{hcm4}) into (\ref{rev2}) and using (\ref{hcm2}) we find
\begin{equation}
T\Gamma^*T|s_j)=\gamma_j|s_j).
\label{hcm8}
\end{equation}
Multiplying (\ref{hcm8}) on the left by $T$, noting that $T^2=1$, and taking the complex conjugate of the resulting equation we find
\begin{equation}
\Gamma T|s_j^*)=\gamma_j^* T|s_j^*).
\label{hcm10}
\end{equation}
Here we have defined the complex-conjugate column vector by
\begin{equation}
(\mu\nu|s_j^*)=\langle\mu|s_j^*|\nu\rangle = [\langle \mu|s_j|\nu\rangle]^*=[(\mu\nu|s_j)]^*.
\label{hcm10a}
\end{equation}
From (\ref{hcm10}) we see that
\begin{equation}
 T|s_j^*)=|s_j^{\dag})
\label{hcm11}
\end{equation}
is an eigenvector of $\Gamma$ with eigenvalue $\gamma_j^*$. With some care in the
case of degeneracies, where several spin modes have the same eigenvalue $\gamma_j$, we can define the modes such that
\begin{align}
T|s_j)&=|s_{j^*}^*)\quad\hbox{or}\quad |s_j^{\dag})=|s_{j^*})\quad\nonumber \\
&\hbox{with}\quad \gamma_j^*=\gamma_{j*}.
\label{hcm12}
\end{align}
Since taking two Hermitian conjugates restores the original matrix, we must have
\begin{equation}
j^{**}=j.
\label{hcm14}
\end{equation}
Using the expressions above we find that an alternate expression for the transposition operator of (\ref{hcm6}) is
\begin{equation}
 T=\sum_j|s_{j^*}^*)\,\lvec s_j|.
\label{hcm18}
\end{equation}
Hopefully, the use of the superscript $*$ to denote the index $j^*$ of the mode $s_{j^*}=s_j^{\dag}$, will not be
confused with the symbol for complex conjugation of a number. The indices for the $n_s=(2I+1)(2S+1)$ modes are the real integers,
$j=0,1,2,\ldots,n_s-1$ and the mode indices $j^*$ are the same real numbers in some permuted order. Both
$s_j$ and its Hermitian conjugate $s_j^{\dag}$ are square matrices in the ground-state subspace of Schroedinger space; $|s_j)$ and $|s_j^{\dag})$ are
column vectors in Liouville space, but $|s_j)^{\dag}=( s_j|$ is a row vector in Liouville space with elements $(s_j|\mu\nu)=(\mu\nu|s_j)^*$.

The matrix elements of the Liouville conjugate $M^{\ddag}=TM^*T$ of a spin-space matrix $M$ are
\begin{align}
&(\mu\nu|M^{\ddag}|\kappa\lambda)=(\nu\mu|M|\lambda\kappa)^*;\nonumber\\
&\lvec s_j|M^{\ddag}|s_k)=\lvec s_{j^*}|M|s_{k^*})^*.
\label{hcm24}
\end{align}

\subsection{Spin-mode expansions \label{spinC}}
For numerical work, it will be convenient to replace the spin-velocity basis vectors $|\mu\nu)\otimes|x_k)$ and $(\mu\nu|\otimes (x_k|$ by
basis vectors with a spin-mode part and a velocity part,
\begin{equation}
|s_j)\otimes|x_k) \quad\hbox{and}\quad \lvec s_j|\otimes (x_k|.
\label{svm2}
\end{equation}
We can expand the density matrix $|\Phi)$ of (\ref{ls22}), on the basis vectors (\ref{svm2}) to find,
\begin{equation}
|\Phi) =\sum_{j}|s_j)\otimes|\chi_j).
\label{svm10}
\end{equation}
The coupled and uncoupled velocity amplitudes of (\ref{svm10}) and (\ref{ls22}) are related by
\begin{equation}
|\chi_j)=\sum_{\mu\nu}\lvec s_j|\mu\nu)|\chi_{\mu\nu}).
\label{svm16}
\end{equation}
%%
%The inverse of (\ref{svm16}) is
%%
%\begin{equation}
%|\chi_{\mu\nu})=\sum_j(\mu\nu|s_j)|\chi_j).
%\label{svm18}
%\end{equation}
%%
%The density matrix (\ref{svm10}) must be Hermitian,
%%
%\begin{eqnarray}
%|\Phi^{\dag}) =\sum_j|s_{j^*})\otimes|\chi_j^*)=|\Phi)=\sum_{j^*}|s_{j^*})\otimes|\chi_{j^*}),\quad
%\label{svm22}
%\end{eqnarray}
%%
%where in analogy to (\ref{hcm10a})
%%

For Hermitian density matrices $\Phi^{\dag}=\Phi$  we must have
\begin{eqnarray}
|\chi_j^*)=|\chi_{j^*}), 
\label{svm24}
\end{eqnarray}
or
\begin{equation}
(x_k|\chi_j^*)=(x_k|\chi_j)^*=(x_k|\chi_{j^*}).
\label{svm23}
\end{equation}
The elements $(x_k|\chi_j)$ of population modes  with $j^*=j$ must be real.
\section{Velocity-Damping \label{velocity}}
We assume that gas-phase collisions cause the density matrix (\ref{ls22}) to change in velocity space at the rate
\begin{equation}
\frac{\partial}{\partial t}|\Phi)=-\gamma_{\rm vd}\sum_{j}|s_j)\otimes A_{\rm vd}|\chi_j).
\label{vev2}
\end{equation}
Here $\gamma_{\rm vd}$ is a characteristic velocity-damping rate, which can be deduced with the aid of (\ref{sd80}) from the spatial diffusion coefficient $D$ and the smallest, non-zero eigenvalue $\alpha_1$
of the collision operator $A_{\rm vd}$. The collision operator is often written in terms of a velocity-changing collision kernel $W_{\rm vd}$ as
\begin{equation}
A_{\rm vd}=1^{\{x\}}-W_{\rm vd}.
\label{vev4}
\end{equation}
The velocity-space unit operator is $1^{\{x\}}=\sum|x_k)(x_k|$.

Atom conservation implies that that the velocity-evolution operators of (\ref{vev4}) must satisfy a constraint analogous to
(\ref{sev10}) for the spin-evolution operators, which we write as
\begin{eqnarray}
\lvec v_0|A_{\rm vd}&=&0,\label{vev8}\\
\lvec v_0|W_{\rm vd}&=&\lvec v_0|.\label{vev6}
\end{eqnarray}
The equilibrium left eigenvector $\lvec v_0|$ is analogous to $\lvec s_0|$ of (\ref{sev12}), and is given by
\begin{equation}
\lvec v_0|=\sum_{k}(x_k|.
\label{vev10}
\end{equation}
The equilibrium right eigenvector $|v_0)$, conjugate to the left eigenvector $\lvec v_0|$ of (\ref{vev10}), is analogous to $|s_0)$ of (\ref{lev8}), and is given by the Maxwellian distribution
\begin{equation}
|v_0)=\frac{\delta x}{\sqrt{\pi}}\sum_{k}|x_k)e^{-x_k^2}.
\label{vev16}
\end{equation}
The analogs of the equilibrium mode constraint (\ref{lev12}) are
\begin{eqnarray}
A_{\rm vd}|v_0)&=&0,\label{vev22}\\
W_{\rm vd}|v_0)&=&|v_0).\label{vev20}
\end{eqnarray}
\subsection{Velocity modes}
In analogy to (\ref{rev2}) and (\ref{lev2}), we assume that the velocity-damping operator $A_{\rm vd}$ has a spectrum of right and left
eigenvectors $|v_n)$ and $\lvec v_n|$ corresponding to the eigenvalue $\alpha_n$ such that
\begin{eqnarray}
W_{\rm vd}|v_n)&=&\varpi_n|v_n)\quad\hbox{and}\quad\lvec v_n|W_{\rm vd}=\lvec v_n|\varpi_n,\label{vev26}\\
A_{\rm vd}|v_n)&=&\alpha_n|v_n)\quad\hbox{and}\quad\lvec v_n|A_{\rm vd}=\lvec v_n|\alpha_n,\label{vev28}
\end{eqnarray}
with
\begin{equation}
\varpi_n=1-\alpha_n.
\label{vev30}
\end{equation}
One of the first clear examples of velocity modes was given by Snider \cite{Snider86}.
There will be the same number $n_x$ of independent eigenvectors $|v_n)$ as there are velocity sample points $x_k$. The eigenvalues are real and non-negative. They can be numbered by the integers $n=0,1,2,\ldots, n_x-1$ such that
\begin{align}
&\varpi_0=1\ge \varpi_1\ge \varpi_2\ge\ldots,\quad\hbox{and}\nonumber \\
&\alpha_0=0\le \alpha_1\le \alpha_2\le\ldots.
\label{vev32}
\end{align}
In analogy to (\ref{lev6}), we assume that the left and right eigenvectors can be chosen to be orthonormal and complete, so that
\begin{equation}
\lvec v_n|v_k)=\delta_{nk},\quad\hbox{and}\quad \sum_n|v_n)\lvec v_n|=1^{\{x\}}.
\label{vev34}
\end{equation}
The values of the summation index are $n=0,1,2,\ldots$
In analogy to (\ref{lev18}) we write
\begin{equation}
W_{\rm vd} =\sum_{n}\varpi_n|v_n)\lvec v_n|,\quad\hbox{and}\quad A_{\rm vd} =\sum_{n}\alpha_n|v_n)\lvec v_n|.
\label{vev36}
\end{equation}
\subsection {Keilson-Storer kernels}
A convenient model kernel $W_a=W_{\rm vd}$ for the velocity-changing collision kernels $W_{\rm vd}$ of (\ref{vev4}) was introduced by Keilson-Storer (KS) \cite{Keilson52}.
In the KS model, it is assumed that a group of atoms, all having the the same sample velocity $x_k$ along the laser beam,
are transformed by an ensemble of single collisions, with various impact parameters and orbital planes, into the distribution of final velocities $x_j$
\begin{equation}
W_{a}(x_j,x_k)=\frac{1}{\delta x}(x_j|W_a|x_k)=\frac{e^{-(x_j-ax_k)^2/b^2}}{b\sqrt{\pi}}.
\label{ks2}
\end{equation}
This newly formed Gaussian distribution is centered at the dimensionless velocity $ax_k$. The ``memory'' parameter $a$ and width $b$ are
\begin{equation}
0\le a<1\quad\hbox{and}\quad b=\sqrt{1-a^2}.
\label{ks3}
\end{equation}
Snider \cite{Snider86} has shown that the KS kernel can be written as the eigenvalue expansion \cite{Morgan10}
\begin{equation}
W_{a}=\sum_n a^n|v_n)\lvec v_n|.
\label{ks4}
\end{equation}
The amplitudes of the left and right eigenvectors can be chosen to be
\begin{equation}
\lvec v_n|x_k)=\frac{H_n(x_k)}{\sqrt{2^n n!}}\quad\hbox{and}\quad (x_k|v_n)=\frac{\delta x H_n(x_k)e^{-x^2_k} }{\sqrt{2^n n!\pi}}.
\label{ks6}
\end{equation}
Here $H_n(x)$ denotes the $n$th Hermite polynomial.
The KS eigenvectors are independent of the memory parameter $a$. Then the KS eigenvalues of
(\ref{vev30}) are
\begin{equation}
\varpi_n=a^n,\quad\hbox{and}\quad \alpha_n=1-a^n.
\label{ks8}
\end{equation}

With their simple analytic form KS kernels (\ref{ks2}) are convenient for numerical work, and they clearly satisfy the normalization constraint (\ref{vev6}) and the Maxwellian constraint (\ref{vev20}). However,
KS kernels with a single memory parameter $a$ do not give very good approximations to kernels inferred from experimental observations, for example, those of Gibble and Gallagher~\cite{Gibble91} or to kernels modeled from realistic interatomic potentials, like those of Ho and Chu~\cite{Ho86}.

Real collisions occur for a large range of impact parameters, or as large numbers of partial waves in a quantum treatment of the scattering. ``Head-on" collisions with small impact parameters will produce large changes in velocity and can be approximately modeled by KS kernels that are close to the strong-collision limit, with $a=0$. ``Grazing-incidence" collisions with large impact parameters will be much more frequent but will produce small changes in velocity. They are better modeled by KS kernels with $a\approx 1$.

With these facts in mind it has often been proposed that a superposition of KS kernels would be a better model, but suggested superpositions have been somewhat inconvenient for numerical modeling. McGuyer {\it et al.}~\cite{McGuyer12} have introduced the ``cusp kernel," a special superposition of KS kernels that is even more convenient for modeling than KS kernels, since cusp kernels and their superpositions can be readily inverted to find steady-state velocity distributions. It is more difficult to invert KS kernels or other collision kernels that have been used in the past. Cusp kernels and their superpositions are also more similar to kernels inferred from experiment \cite{Gibble91} or from realistic interatomic potentials \cite{Ho86}.

\subsection{Cusp kernels}
Cusp kernels describe an ensemble of collisions, with each sample collision described with a KS kernel of memory parameter $a$.
The probability to find the the memory parameter between $a$ and $a+da$ is $P_s(a)da$, where the probability density is
\begin{equation}
P_s(a)=sa^{s-1}.
\label{ck2}
\end{equation}
We will call the parameter $s$ the ``sharpness," and the corresponding velocity-changing collision kernel of (\ref{vev4}) will be denoted by $W_{\rm vd}=C_s$. The sharpness $s$ will be a real, positive number for velocity-changing collision kernels $W_{\rm vd}$, but for ``resolvent kernels" $\overline W$, which we will discuss below, the sharpness can have an imaginary part. An explicit expression for the cusp kernel is
\begin{equation}
C_s=\int_0^1 W_aP_s(a)da=\sum_n\frac{s|v_n)\lvec v_n|}{s+n}.
\label{ck6}
\end{equation}
The eigenvalues of (\ref{vev30}) are
\begin{equation}
\varpi_n=\frac{s}{s+n} \quad\hbox{and}\quad \alpha_n=\frac{n}{s+n}.
\label{ck7}
\end{equation}
In the limit $n\to \infty$ we have
\begin{equation}
\varpi_{\infty}=0 \quad\hbox{and}\quad \alpha_{\infty}=1,
\label{vev33}
\end{equation}
for both cusp kernels (\ref{ck7}) and KS kernels (\ref{ks8}).

\begin{figure}[ht]
\includegraphics[width=7.8cm]{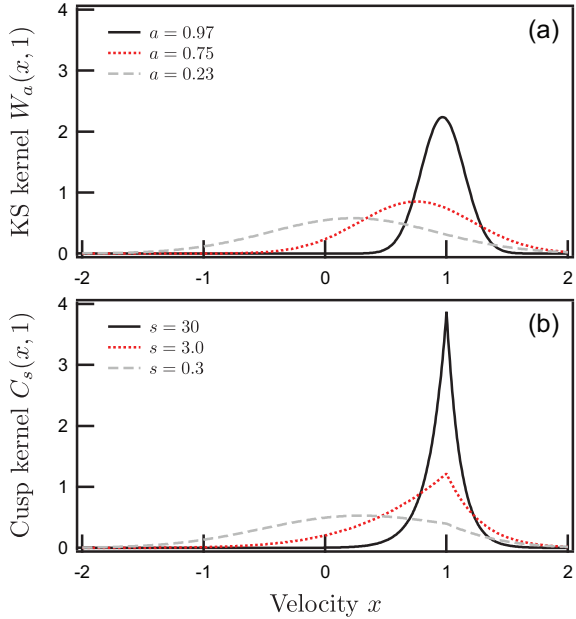}
\caption{(Color online) Keilson-Storer kernels $W_a$ of (\ref{ks2}) with representative memory parameters $a$, compared to cusp kernels $C_s$ of (\ref{ck8}) with sharpness parameters $s$ that have the same
expectation value, $\langle a\rangle = \int_0^1 P_s(a)a da = s/(s+1)$, of the memory parameter. As $a\to 0$ and $s\to 0$, the KS and cusp kernels generate Maxwellian distributions.
 \label{Fig1_McGuyer_PRL}}
\end{figure}

Morgan and Happer \cite{Morgan10} have shown that the matrix elements of (\ref{ck6}) can be summed to give
\begin{equation}
C_s(x_j,x_k)
=\frac{s2^s\Gamma(s)}{\sqrt{\pi}}e^{{x_k}^2} R_s(-x_<)R_s(x_>).
\label{ck8}
\end{equation}
Here $\Gamma(s)$ is the Euler gamma function, $x_>$ is the greater of the two variables $x_j$ and $x_k$, and $x_<$ is the lesser. The ``right function" $R_s(z)$ can be represented with the power series \cite{Morgan10}, convergent for all finite $z$,
\begin{equation}
R_s(z)=\sum_{n=0}^{\infty}\frac{\sqrt{\pi}(-z)^n}{n!2^{s-n}\Gamma(\frac{1}{2}+\frac{s}{2}-\frac{n}{2})}.
\label{ck10}
\end{equation}
For sufficiently large sharpness, $|s|\gg 1$, one can evaluate the cusp kernel (\ref{ck8}) with the asymptotic expression
\begin{equation}
2\ln C_s(x_j,x_k)=x_k^2-x_j^2-2|x_j-x_k|\sqrt{2s}+\ln\left(\frac{s}{2}\right).
\label{ck12}
\end{equation}
A comparison of KS kernels and cusp kernels is shown in Fig.~\ref{Fig1_McGuyer_PRL}.

\begin{figure}[ht!]
\includegraphics[width=7.8cm]{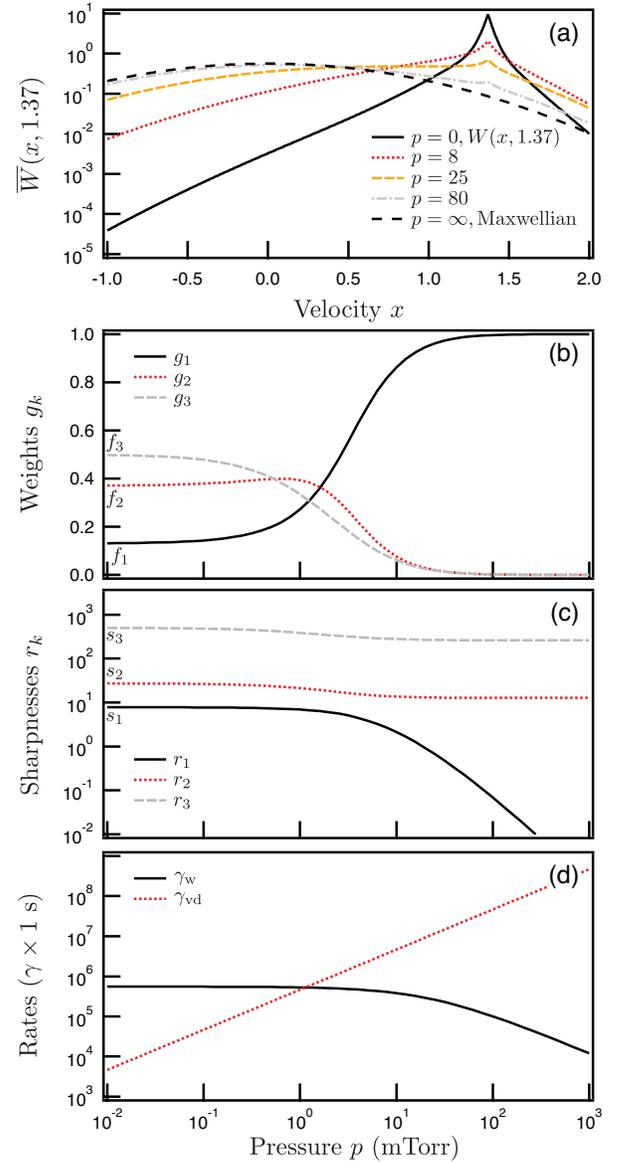}
\caption{(Color online) Resolvent kernels $\overline W$ versus buffer-gas pressure $p$ for the initial velocity $y=1.37$. Here, the collision kernel $W_{\rm vd}$ of (\ref{ck14}) is a three-cusp fit from Ref.~\cite{McGuyer12} to a measured kernel for Rb in He from Ref.~\cite{Gibble91}. The superposition parameters are $f_j = [0.13, 0.37, 0.50]$ and $s_j=[7.8,27.2,500]$. (a) As described by the expressions of Section \ref{mckernel}, with increasing buffer-gas pressure the resolvent kernel $\overline{W}$ evolves from a low-pressure, single-collision limit, $\overline{W}=W_{\rm vd}$, to a high-pressure limit, which is a Maxwellian distribution $\overline{W}=|v_0)\lvec v_0|$. (b) With increasing pressure the weights $g_k$ of the resolvent kernel tend to the limit $g_k = [1,0,0]$. (c) The sharpnesses $r_k$ of the resolvent kernel diminish with pressure, and in the high pressure limit, $r_1 \rightarrow 0$, give a Maxwellian distribution for the only resolvent cusp with appreciable weight, $g_1\approx 1$. (d) The pressure-dependent rates $\gamma_{\rm vd}$ of (\ref{sd80}) and $\gamma_{\rm w}$ of (\ref{pdr12}) are for $a=1$ mm, $b=5$ mm, $T = 60~^\circ$C, and $D_0 = 0.15$~cm$^2$\,s$^{-1}$ at 760 Torr. The transition from low-pressure to high-pressure conditions occurs when $\gamma_{\rm w}\approx \gamma_{\rm vd}$.
\label{Fig2}}
\end{figure}

As shown by McGuyer {\it et al.}~\cite{McGuyer12}, one can fit collision kernels inferred from experimental measurements very well with the superposition of a few ($m\approx 3$) cusp kernels: 
\begin{equation}
W_{\rm vd} = \sum_{k+1}^mf_k C_{s_k} \quad\hbox{with}\quad \sum_{k=1}^m f_k=1.
\label{ck14}
\end{equation}
Each cusp of sharpness $s_k$ makes a fractional contribution $f_k>0$ to the overall damping operator $W_{\rm vd}$.
An example of a multicusp collision kernel from McGuyer {\it et al.} \cite{McGuyer12} is shown as the curve labeled $p=0$ in Fig.~\ref{Fig2}a. This is the kernel that
was used for the representative models of experimental data in Figs. \ref{Fig5}, \ref{Fig6} and \ref{Fig8}. The other parts of Fig.~\ref{Fig2} are discussed in Section~\ref{mckernel} where we show how to find multicusp resolvent kernels from multicusp collision kernels.

\section{Wall-Damping \label{pressure}}
Most existing data on velocity-selective optical pumping has come from laboratory experiments, where the atoms either experience no buffer gas collisions at all, or they collide with buffer gases like He, Ne, Ar, Kr, Xe, N$_2$, or H$_2$, where the rate $\gamma_{\rm sd}$ of spin-changing collisions in (\ref{sev4}) is orders of magnitude smaller than the rate $\gamma_{\rm vd}$ of velocity changing collisions in (\ref{vev2}). Under these conditions, the damping rates $\gamma_j$ of the spin modes can be well approximated by $\gamma_j=0$ for population modes
with $j^*=j$, and by $\gamma_j=i\omega_j$ for coherence modes with $j^*\ne j$ and Bohr frequency $\omega_j=-\omega_{j^*}$. Under these laboratory conditions, the effects of the buffer gas on velocity-selective optical pumping are characterized by the rate $\gamma_{\rm vd}$ of velocity-changing collisions and by the effective rate $\gamma_{\rm w}$ of collisions with the wall. Walls with no special coatings destroy the spin polarization of impinging atoms and release unpolarized atoms with a Mawellian distribution of velocities.

For analyzing the physics of Na guidestar atoms there is no need to consider wall collisions, but the O$_2$ molecules and O atoms of the buffer gas at an altitude of 90--100~km have rates $\gamma_{\rm sd}$ for spin-changing collisions that are comparable to or larger than the rates $\gamma_{\rm vd}$
for velocity-changing collisions. Under these conditions, the damping rates $\gamma_j$ of the spin modes have relatively large real parts due to spin-changing collisions.

\subsection{Walls}
In analogy to (\ref{sev2}) and (\ref{vev2}) we write wall relaxation as
\begin{equation}
\frac{\partial}{\partial t}|\Phi)=-\gamma_{\rm w} A_{\rm w}|\Phi).
\label{ws2}
\end{equation}
The effective collision rate of atoms with the walls, $\gamma_{\rm w}$, is a real, positive number.
We assume that every atom that hits a wall sticks, and is replaced by a different atom that evaporates with no spin polarization and with a Maxwell distribution of velocities. The wall depolarization operator is simply
\begin{equation}
A_{\rm w}=1-|\text{\o})\lvec\text{\o}|.
\label{ws4}
\end{equation}
Here we use (\ref{sev12}) with (\ref{vev10}) to write the equilibrium spin-velocity row vector for the full spin-velocity space as
\begin{equation}
\lvec \text{\o}|=\lvec s_0|\otimes\lvec v_0|.
\label{ws6}
\end{equation}
The corresponding equilibrium column vector,
\begin{equation}
|\text{\o})=|s_0)\otimes| v_0),
\label{ws8}
\end{equation}
is the density matrix for atoms with no spin polarization and a Maxwell distribution of velocities. For the modifications needed to account for walls that only partially depolarize the spins and lead to shifts of the coherence frequencies, see the work of Wu {\it et al.}~\cite{Wu88}.

\subsection{Cylindrical cells}
 As a semiquantitative example of how to estimate wall relaxation rates $\gamma_{\rm w}$, let us suppose that a pump laser fills a cylindrical volume of radius $a$ on the axis of a cylindrical cell of radius $b>a$. Let $\chi$ be some spin-mode amplitude of
 the density matrix that is pumped at a rate $S$ inside the pump laser beam. We assume axial symmetry so that $\chi=\chi(r)$ depends only on the distance $r$ from the cylinder axis. Then the diffusion equation is
\begin{equation}
\frac{\partial}{\partial t}\chi-D\nabla^2 \chi= \left \{\begin{array}{ll} S\quad&\mbox{for\quad$r\le a$}\\
 0 &\mbox{for\quad$a< r\le b$}\end{array}\right..
\label{pdr2}
\end{equation}
The spatial diffusion coefficient can be measured experimentally and the results are often expressed as
\begin{equation}
D=D_{0}\frac{p_0}{p},
\label{pdr4}
\end{equation}
where $p$ is the gas pressure and $D_0$ is the diffusion coefficient at the reference pressure $p_0$, which is usually 1~atm.

One can integrate the steady-state version of (\ref{pdr2}) with the boundary condition $\chi(b)=0$ and with $\chi$ and $d\chi/dr$ continuous at $r=a$ to find the solution
\begin{equation}
\frac{\chi}{S}= \frac{(a^2-r^2)}{4D}+\frac{a^2 }{2D}\ln\left(\frac{b}{a}\right ),\quad\hbox{if}\quad r\le a,
\label{pdr6}
\end{equation}
or
\begin{equation}
\frac{\chi}{S}= \frac{a^2}{2D}\ln\left(\frac{b}{r}\right ),\quad\hbox{if}\quad a<r\le b.
\label{pdr8}
\end{equation}
For pressures high enough for the diffusion equation to be valid, we define the mean spin polarization $\overline\chi$ sampled by the probe beam, and the mean time $\tau_{\rm w}$ for an atom to be lost from the probe beam by
\begin{equation}
\overline{\chi}= \frac{1}{\pi a^2}\int_0^a \chi(r)2\pi r dr=S\tau_{\rm w}.
\label{pdr10}
\end{equation}
Substituting (\ref{pdr6}) into (\ref{pdr10}) and adding a representative free flight time, $a/v_D$, for the atom to escape the pump beam in the limit of very low pressures, we find
\begin{equation}
\tau_{\rm w}=\frac{1}{\gamma_{\rm w}}=\frac{a}{v_D}+\frac{a^2}{8D}\left[1+4\ln\left(\frac{b}{a}\right)\right].
\label{pdr12}
\end{equation}
\section{Evolution in the Dark \label{dark}}
Summing the rates of change of the density matrix (\ref{svm10}), from wall collisions (\ref{ws2}), from hyperfine interactions and spin-relaxing collisions (\ref{sev2}), with $\Gamma$ given by (\ref{lev18}), and from velocity relaxing collisions (\ref{vev2}), with $A_{\rm vd}$ given by (\ref{vev36}), we find that the evolution rate from all sources
except optical pumping is
\begin{align}
\frac{\partial}{\partial t}|\Phi)&=
-\sum_j(\gamma_{\rm w}+\gamma_j)|s_j)\otimes|\chi_j)\nonumber \\
&+\gamma_{\rm vd}|s_j)\otimes A_{\rm vd}|\chi_j)+\gamma_{\rm w}|{\text{\o}})\lvec v_0|\chi_0).
\label{rid12}
\end{align}
For laboratory and guidestar experiments with weak pumping light, the unpolarized part of the density matrix will have a Maxwellian distribution
\begin{equation}
|\chi_0)=| v_0).
\label{rid14}
\end{equation}
For spin-polarized parts of the density matrix with $j\ne 0$, we can multiply (\ref{rid12}) on the left by $\lvec s_j|\otimes 1^{\{x\}}$ to find
\begin{equation}
\frac{\partial}{\partial t}|\chi_j)=-(\gamma_{\rm w}+\gamma_j+\gamma_{\rm vd}A_{\rm vd})|\chi_j)=-K_j|\chi_j).
\label{rid16}
\end{equation}
Here we have introduced a damping kernel for the $j$th spin mode,
\begin{equation}
 K_j=\sum_n\gamma_{jn}|v_n)\lvec v_n|.
\label{rid18}
\end{equation}
The kernel $K_j$ includes the evolution of the spin due to gas-phase and wall collisions, hyperfine interactions, and precession in an external magnetic field. Also included in $K_j$ are velocity-changing collisions.
The characteristic relaxation rate for the $n$th velocity mode of the spin mode $j$ is
\begin{equation}
 \gamma_{jn}=\gamma_{\rm w}+\gamma_j +\alpha_n\gamma_{\rm vd}.
\label{rid20}
\end{equation}
 Here $\gamma_{\rm w}$ is the effective collision rate of polarized atoms with the wall, $\gamma_j$ is the sum of relaxation due to gas-phase collisions ($=0$ for most laboratory experiments) plus a factor $i\omega_j$ for the Bohr frequency $\omega_j$ of the spin mode $|s_j)$, and $\alpha_n\gamma_{\rm vd}$ is the relaxation rate of the velocity mode $|v_n)$ due to gas-phase collisions at the rate $\gamma_{\rm vd}$.
According to (\ref{hcm12}) and (\ref{rid20}) the coherence damping rates can be complex, but they must satisfy the identity
\begin{equation}
\gamma^*_{jn}=\gamma_{j^*n}.
\label{rid24}
\end{equation}
A special case of (\ref{rid20}) is $\gamma_{00}=0$.
\section{Optical Pumping\label{pumping}}
\begin{figure}[t]
\includegraphics[width=\columnwidth]{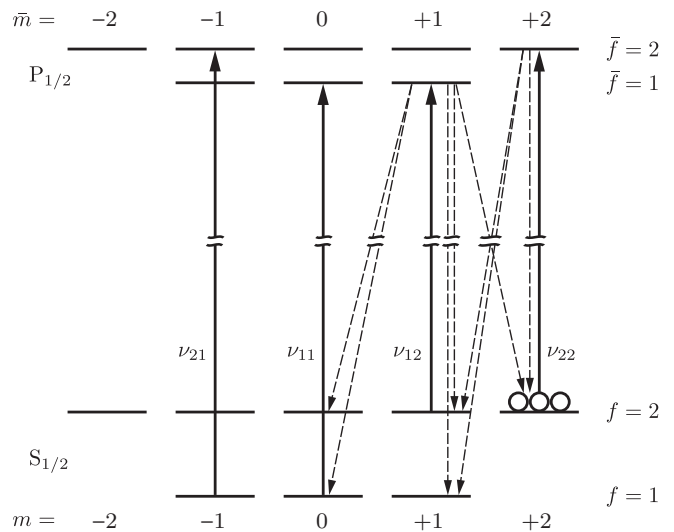}
\caption{Energy sublevels of an alkali-metal atom, resonant frequencies $\nu_{\bar f f}$, and representative transitions for absorption of light (depopulation pumping), or spontaneous emission of light (repopulation pumping). In this example the pumping light is linearly polarized parallel to a small, externally applied magnetic field. Because of the Doppler shift, the resonance frequencies depend on the atomic velocity and on the direction of propagation of the light.
\label{Fig3}}
\end{figure}

The basic processes involved in optical pumping are sketched in Fig.~\ref{Fig3}.
We consider monochromatic pumping light of temporal frequency $\omega=2\pi\nu$ and spatial frequency $k=\omega/c$, propagating along the unit vector $\bzeta$. We assume that the light
intensity is large enough to cause substantial spin polarization of the ground-state atoms, but that it is not so intense that it produces substantial population of the excited state. Most laboratory experiments on velocity-selective optical pumping, and most guidestar systems are in this regime. It is a straightforward extension to account for the saturation of the optical pumping of the ground state from intense, repetitively-pulsed lasers. For laboratory experiments, the atoms will also be subject to a weak, counter-propagating probe beam, usually a small fraction of light from the
source of the pump beam. As the laser frequency changes, resonant changes in the
attenuation of the probe beam occur for laser frequencies where the velocities of atoms spin-polarized by the pump beam also have the right Doppler
shift to resonantly absorb light from the retro-reflected probe beam.
The classical electric field of a monochromatic light beam for an atom located at the position ${\bf r}$ at time $t$ is
\begin{equation}
{\bf E}=\tilde{{\bf E}}e^{i(\sigma k \bzeta\cdot {\bf r}-\omega t)}+\tilde{{\bf E}}^*e^{-i(\sigma k \bzeta\cdot {\bf r}-\omega t)}
\label{opm2}
\end{equation}
The direction index $\sigma=\pm 1$ will be taken to be $\sigma=1$ for the pumping beam. For a probe beam that propagates in the same direction as the pumping beam we will also have $\sigma=1$, and for for a counter-propagating probe beam we will have $\sigma = -1$. The amplitude $\tilde{\bf E}$ of the probe beam may have a different polarization that than the pump beam.
The optical intensity $I$ (in units of erg\,cm$^{-2}$\,s$^{-1}$) is given in terms of the field amplitude $\tilde{\bf E}$ of (\ref{opm2}) by
\begin{equation}
I=\frac{c}{2\pi}|\tilde{{\bf E}}|^2.
\label{opm4}
\end{equation}
As discussed in connection with (5.52) of OPA \cite{OPA}, the interaction of the light with the atoms can be well approximated by the sum of conjugate interaction matrices $\tilde{\mathcal{V}}+\tilde{\mathcal{V}}^{\dag}$, where
\begin{equation}
\tilde{\mathcal{V}}=-\tilde{\bf D}^{\dag}\cdot \tilde{{\bf E}} \quad\hbox{and}\quad \tilde{\mathcal{V}}^{\dag}=-\tilde{\bf D}\cdot \tilde{{\bf E}}^*.
\label{opm9}
\end{equation}
The component $\tilde{\bf D}^{\dag}$ of the electric dipole moment operator ${\bf D}$, when operating on a ground state basis $|\mu\rangle$, produces an excited-state
basis $|\bar\mu\rangle$, and vice versa for $\tilde{\bf D}$.
As shown in connection with (6.7) pf OPA \cite{OPA}, the unsaturated optical coherence generated between excited-state and ground-state sublevels of the atom is proportional
to the matrix,
\begin{equation}
\tilde{\mathcal{W}}=\tilde{\mathcal{V}}./\mathcal{E}^{\{eg\}}.
\label{opm10}
\end{equation}
In (\ref{opm10}) we use the ``dot-slash" symbol ($./$) to denote element-by-element division of the matrices with the same dimensions.

\emph{Resonant velocities.}\quad
From (5.84), (5.88), and (5.92) of OPA \cite{OPA} we see that the energy-denominator matrix that occurs in (\ref{opm10}) is
\begin{equation}
\mathcal{E}^{\{eg\}}_{\bar\mu\mu}= \hbar(\sigma k^{\{eg\}}v_k-\omega +\omega_{\bar\mu\mu})-i\hbar \gamma_{\rm c}=\hbar\gamma_D
\left(\sigma x_k- z_{\bar\mu \mu}\right).
\label{rv2}
\end{equation}
The $k$th discretized component of the atomic velocity along the pump beam is $v_k={\bf v}\cdot \bzeta=x_kv_D$. The photon momentum is very nearly
$\sigma \hbar k^{\{eg\}} \bzeta $, where the spatial frequency is $k^{\{eg\}}$.
The complex resonant velocities are determined by the optical frequency $\omega$ and by the optical Bohr frequencies $\omega_{\bar\mu\mu}$ of the atom, and given by
\begin{equation}
z_{\bar \mu \mu}=x_{\bar \mu \mu}+iy_{\rm c}.
\label{rv5}
\end{equation}
The real and imaginary parts of the resonant velocities are
\begin{equation}
x_{\bar \mu \mu}=\frac{\omega-\omega_{\bar\mu\mu}} {\gamma_D} \quad\hbox{and}\quad y_c=\frac{\gamma_{c}}{\gamma_D},
\label{rv6}
\end{equation}
with $\gamma_D=v_D k^{\{eg\}}$.
We assume that the damping rates associated with the optical coherences $\bar \mu\mu$ are the same and given by
\begin{equation}
\gamma_{c}=\frac{1}{2\tau^{\{e\}}}+\gamma_{\rm oc},
\label{rv8}
\end{equation}
where $\tau^{\{e\}}$ is the natural radiative lifetime of the excited atom. The collisional damping rate $\gamma_{\rm oc}$ of the optical coherence is nearly negligible for most
VSOP situations. In modeling, the parameter $\gamma_{\rm oc}$ will be used to approximately account for the frequency linewidth of the laser and for slight misalignment of the pump
and retro-reflected probe beam.

\begin{figure*}[t]
\includegraphics[width=\textwidth]{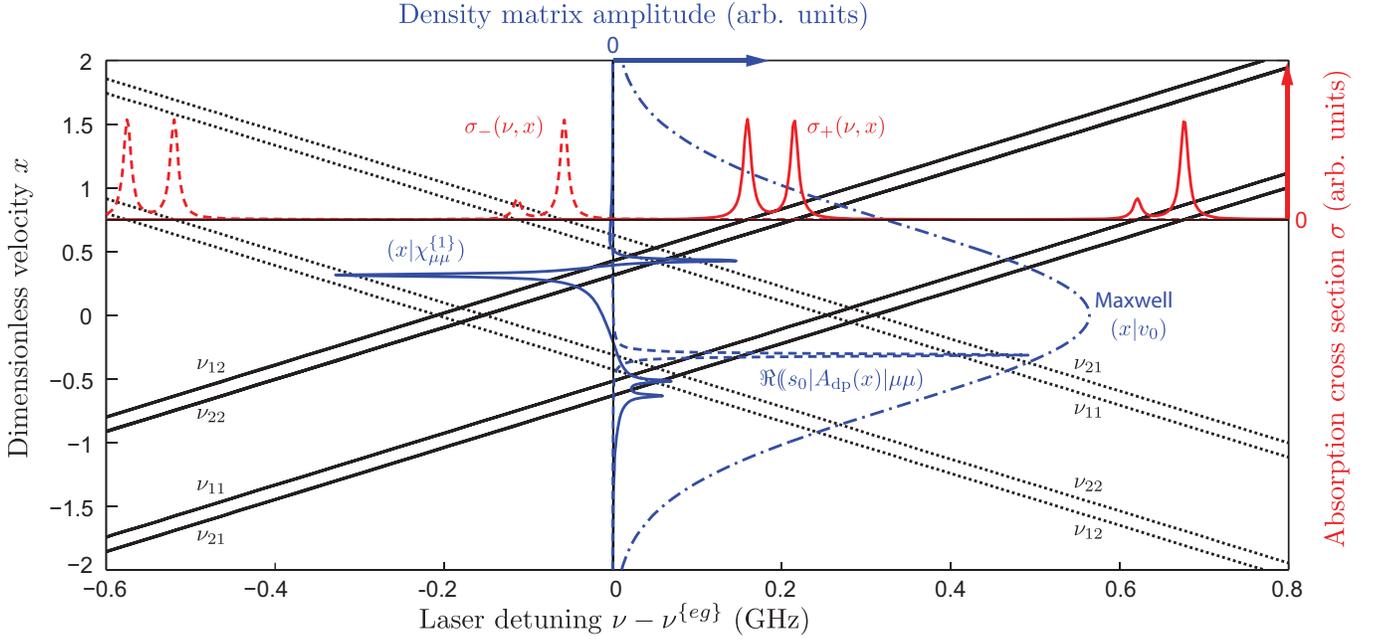}
\caption{(Color online) Resonant velocities $\pm x_{\bar\mu\mu}$ of (\ref{rv6}) for atoms moving along a pump beam (the solid lines sloping up to the right) and for a retroreflected and attenuated probe beam (the dotted lines sloping downward to the right) as a function of the detuning of the laser frequency from the center of the 770~nm D1 line of $^{39}$K atoms in a magnetic field of $B=1$~G. Both the pump and probe light are linearly polarized parallel to the magnetic field, as indicated in Fig.~\ref {Fig3}. The absorption frequencies from Fig.~\ref{Fig3} to which each group of resonant velocities correspond are denoted by $\nu_{\bar f f}$. For a group of unpolarized atoms with dimensionless velocity $x=0.75$, we display the absorption cross sections $\sigma_+(\nu,x)$ of a pump beam (the solid line), and probe beam $\sigma_{-}(\nu,x)$ (the dashed line). Plotted as solid lines along the vertical axis at the center figure are the first-order population shifts $(x|\chi^{\{1\}}_{\mu\mu})$ of (\ref{sr12}) as a function of velocity $x$ for the ground-state sublevel $|\mu\rangle$ with low-field quantum numbers $f=2$ and $m=2$, produced by a pump beam with a frequency detuning of 0~GHz. Plotted as a dashed line on the same scale is the velocity dependence of the factor $\Re\lvec s_0| A_{\rm dp}(x)|\mu\mu)$ of (\ref{dpp18}) for the probe beam.
Collisions in a 10~mTorr buffer gas at $T=50~^\circ$C have transferred atoms in the sublevel $|\mu\rangle$ from their four resonant excitation velocities to other velocities, some of which are in resonance with the probe beam. For these curves, $a=0.1$~cm, $b=0.5$~cm, and $D_0=0.1$~cm$^2$\,s$^{-1}$. The dot-dash line is a Maxwellian distribution of velocities, $(x|v_0)$ of (\ref{vev16}). Strong VSOP resonances may be observed at laser frequencies that make pump and probe resonant velocities equal, within the envelope of the Maxwellian velocity distribution.
\label{Fig4}}
\end{figure*}

The pumping will be resonantly enhanced for velocities $x_k$ that are close to the resonant velocities $x_{\bar\mu\mu}$, that is, for $x_k\approx x_{\bar\mu\mu}$
Similarly, the probe absorption will be resonantly enhanced for velocities as close as possible to the resonant velocities $\sigma x_{\bar\nu\nu}$, that is, for
$x_k\approx\sigma x_{\bar\nu\nu}$. If the probe beam propagates parallel to the pump beam ($\sigma=1$), every resonant velocity for the pump beam will also be resonant for the probe beam. However, if the probe beam is counter-propagating with $\sigma =-1$, the resonant conditions can be satisfied for the same velocity group only if the laser detuning is such that two different optical transitions, $\bar\mu\mu$ and $\bar\nu\nu$, have equal and opposite resonant velocities,
\begin{equation}
x_{\bar\mu\mu}=-x_{\bar\nu\nu}.
\label{rv14}
\end{equation}
This is the condition for saturated-absorption resonances with counter-propagating laser beams. In Fig.~\ref{Fig4} we show how the resonant velocities $x_{\bar\mu\mu}$ and their negatives $-x_{\bar\nu\nu}$ depend on laser detuning for a $^{39}$K atom in a magnetic field of $1$~G.

\emph{Residue matrices.}\quad Using (\ref{rv2}) we can write (\ref{opm10}) as
\begin{equation}
\tilde{\mathcal{W}}(x_k)=\sum_{\bar \mu \mu}\frac{\tilde{\mathcal{W}}^{\{\bar \mu \mu\}}}{\sigma x_k- z_{\bar \mu \mu}}.
\label{rv16}
\end{equation}
For each optical transition between a ground-state sublevel $\mu$ and and excited-state sublevel $\bar\mu$, we have defined a $g^{\{e\}}\times g^{\{g\}}$ ``residue matrix" with a single nonzero element
\begin{equation}
\langle\bar\nu|\tilde{\mathcal{W}}^{\{\bar \mu \mu\}}|\nu\rangle =\tilde{\mathcal{W}}^{\{\bar \mu \mu\}}_{\bar\nu\nu}=
\delta_{\bar\nu \bar \mu}\delta_{\nu\mu}\frac{\tilde{\mathcal{V}}_{\bar \mu \mu}}{\hbar \gamma_D}.
\label{rv18}
\end{equation}
The residue matrices are independent of the laser frequency $\omega$ and the atomic velocity $x_k$. They give the relative contributions to the pumping rate
of the optical transitions from sublevel $|\mu\rangle$ to $|\bar\mu\rangle $.
\subsection{\bf Depopulation pumping.}
A natural rate to characterize velocity-dependent optical pumping is the maximum possible pumping rate,
\begin{equation}
\gamma_{\rm mx}=\frac{I}{\hbar\omega^{\{eg\}}}\sigma_{\rm mx},
\label{dpp2}
\end{equation}
for a hypothetical isotope with no externally applied magnetic field, with vanishing hyperfine coupling coefficients, with no spin polarization, and with zero velocity.
Such hypothetical atoms would have only a single optical Bohr frequency $\omega_{\bar\mu\mu}=\omega^{\{eg\}}$, and their absorption cross section for light of frequency $\omega$ would be independent of optical polarization and given by
\begin{equation}
\sigma_{\rm eq}=\frac{\gamma_c^2}{(\omega-\omega^{\{eg\}})^2+\gamma_c^2}\,\sigma_{\rm mx},
\label{dpp4}
\end{equation}
where the maximum cross section is
\begin{equation}
\sigma_{\rm mx}=\frac{2\pi r_e c f^{\{ge\}}}{\gamma_c}.
\label{dpp6}
\end{equation}
Here $f^{\{ge\}}$ is the oscillator strength, $r_e$ is the classical electron radius, $c$ is the speed of light, and the optical coherence damping rate was given by (\ref{rv8}).

We will write the evolution of the mode amplitude $|\chi_j)$ due to depopulation pumping as
\begin{equation}
\frac{\partial}{\partial t}(x_k|\chi_j)=-\gamma_{\rm mx}\sum_{l}\lvec s_j|{A}_{\rm dp}(x_k)|s_l)(x_k|\chi_l).
\label{dpp8a}
\end{equation}
We have neglected the small changes in velocity $x_k$ caused by absorption and emission of light. The depopulation pumping operator $A_{\rm dp}$ can be written as
\begin{equation}
{A}_{\rm dp}(x_k)= \tilde{A}_{\rm dp}(x_k)+\tilde{A}_{\rm dp}^{\ddag}(x_k).
\label{dpp10}
\end{equation}
The Liouville-conjugate matrix $\tilde{A}_{\rm dp}^{\ddag}$ is defined in terms of $\tilde{A}_{\rm dp}$ by (\ref{hcm4}). We can use (6.12) of OPA \cite{OPA} and (\ref{rv16}) to find
\begin{equation}
\tilde{A}_{\rm dp}(x_k)
=\sum_{\bar \mu\mu}\frac{\tilde R_{\rm dp}^{\{\bar \mu \mu\}}}{\sigma x_k- z_{\bar\mu\mu}},
\label{dpp12}
\end{equation}
where the residue matrix is given by
\begin{equation}
\tilde R_{\rm dp}^{\{\bar \mu \mu\}}=\frac{1}{i\hbar\gamma_{\rm mx}}\left[\tilde{\mathcal{V}}^{\dag}\tilde{\mathcal{W}}^{\{\bar \mu \mu\}}\right]^{\flat}.
\label{dpp14}
\end{equation}
As discussed in (4.4.1) of OPA \cite{OPA}, a $g^{\{g\}}\times g^{\{g\}}$  Schr\"odinger-space matrix $M$ is transformed into into a corresponding $g^{\{g\}2}\times g^{\{g\}2}$ ``flat matrix" $M^{\flat}$ of Liouville space by the Kronecker product
\begin{equation}
M^{\flat}= 1^{\{g\}}\otimes M.
\label{dpp13}
\end{equation}
\subsection{Absorption rates and optical cross sections}
The rate of depletion of ground-state atoms by depopulation pumping is
\begin{equation}
-\sum_k\frac{\partial}{\partial t}(x_k|\chi_0)=\langle \delta \Gamma\rangle=\frac{I\langle\sigma\rangle}{\hbar\omega^{\{eg\}}},
\label{dpp16}
\end{equation}
where $\langle \delta \Gamma\rangle$ is the rate at which spin-polarized atoms absorb or scatter light for a light beam of intensity $I$, and
the corresponding absorption cross section of the spin polarized atoms is $\langle \sigma\rangle $. Combining (\ref{dpp16}) with (\ref{dpp8a}) we find
\begin{equation}
\langle \delta \Gamma\rangle= \langle \delta \tilde\Gamma\rangle+\langle \delta \tilde\Gamma\rangle^*.
\label{dpp17}
\end{equation}
where
\begin{equation}
\langle \delta \tilde\Gamma\rangle =\gamma_{\rm mx}\sum_{jk}\lvec s_0|\tilde A_{\rm dp}(x_k)|s_j)(x_k|\chi_j).
\label{dpp18}
\end{equation}

\emph{Mean pumping rate.} We define the mean optical pumping rate, $\gamma_{\rm op}$ of the atoms
as the value of the pumping rate $\langle \delta \Gamma\rangle$ for unpolarized atoms with a Maxwellian distribution of velocities, and therefore
with the spin mode amplitudes

\begin{equation}
(x_k|\chi_j)=(x_k|v_0)\delta_{j0},
\label{dpp19a}
\end{equation}
Then we can use (\ref{dpp19a}) with (\ref{dpp18}) and (\ref{dpp12}) to write the amplitude of the mean pumping rate as
\begin{eqnarray}
\tilde\gamma_{\rm op}&=&\gamma_{\rm mx}\sum_{\bar\mu\mu k}\frac{\lvec s_0|\tilde R^{\{\bar\mu\mu\}}_{\rm dp}|s_0)(x_k|v_0)}{\pm x_k-z_{\bar\mu\mu}}\nonumber\\
&=&i\sqrt{\pi}\gamma_{\rm mx}\sum_{\bar\mu\mu}\lvec s_0|\tilde R^{\{\bar\mu\mu\}}_{\rm dp}|s_0) w(z_{\bar\mu\mu}).
\label{dpp22}
\end{eqnarray}
The integral over the velocity distribution (that is, the sum on $k$ in (\ref{dpp22})) can be written in terms of the Faddeeva function \cite{Abramowitz}, which for $ \Im (z)\ge 0$ is given by
\begin{equation}
w(z) =\frac{1}{i\pi}\int_{-\infty}^{\infty}\frac {e^{-t^2} dt}{t-z}.
\label{fad2}
\end{equation}
The Faddeeva function is a superposition of Lorentzians with a Gaussian distribution of resonance frequencies. It is often called a Voigt profile, and it can be evaluated with a very efficient computer algorithm due to Weideman \cite{Weideman94}.

The mean pumping rate $\gamma_{\rm op}$ of (\ref{dpp22}) depends on the laser frequency through the parameter $z_{\bar\mu \mu}$ and is proportional to the laser intensity
because of the factor $\gamma_{\rm mx}$. For large magnetic fields, $\gamma_{\rm op}$ also depends on the laser polarization. The equilibrium absorption cross section for unpolarized atoms with a Maxwellian distribution of velocities to absorb light is
\begin{equation}
\sigma_{\rm op}=\frac{\hbar\omega^{\{eg\}} \gamma_{\rm op}}{I},
\label{opm24}
\end{equation}
where the light intensity $I$ was given by (\ref{opm4}).

\emph{Repopulation pumping.} \quad
In analogy to (\ref{dpp8a}), the rate of change of the spin-mode amplitudes due to repopulation pumping is
\begin{equation}
\frac{\partial}{\partial t}(x_k|\chi_j)=\gamma_{\rm mx}\sum_{l}\lvec s_j|A_{\rm rp}(x_k)|s_l)(x_k|\chi_l).
\label{rpp2}
\end{equation}
In analogy to (\ref{dpp12}) the amplitude of the repopulation pumping operator is
\begin{equation}
\tilde A_{\rm rp}(x_k) =\sum_{\bar \mu\mu}\frac{\tilde R_{\rm rp}^{\{\bar \mu \mu\}}}{\sigma x_k-z_{\bar\mu\mu}}.
\label{rpp6}
\end{equation}
Using (6.71) and (6.35) of OPA \cite{OPA}, we find
\begin{equation}
\tilde R_{\rm rp}^{\{\bar\mu\mu\}}=\frac{ A^{\{ge\}}_{\rm s}(1+i\tau^{\{e\}} H^{\{e\}}/\hbar)^{\odot\, -1}\tilde{\mathcal{V}}^{*}\otimes
\tilde{\mathcal{W}}^{\{\bar\mu\mu\}}}{i\hbar\gamma_{\rm mx}}.
\label{rpp8}
\end{equation}
Here $H^{\{e\}}$ excited-state Hamiltonian, and $\tau^{\{e\}}$ is the radiative lifetime of the excited atoms.
As defined by (4.91) of OPA \cite{OPA}, only the diagonal elements of the o-dot transform $X^{\odot}$ of a Schr\"odinger-space matrix $X$ are nonzero,
\begin{align}
(\mu\nu|X^{\odot}|\mu\nu)&=\langle \mu|X|\nu\rangle, \quad\hbox{and}\nonumber \\
\quad(\mu\nu|X^{\odot -1}|\mu\nu)&=\langle \mu|X|\nu\rangle^{-1}.
\label{rpp10}
\end{align}
The spontaneous emission matrix is given by (5.50) of OPA \cite{OPA} as
\begin{equation}
A_{\rm s}^{\{ge\}}=\frac{(2J+1)}{3|\mathcal{D}|^2}\sum_k \tilde D_k^{*}\otimes \tilde D_k.
\label{rpp12}
\end{equation}
The sum extends over the projections $\tilde D_k={\bf x}_k\cdot\tilde {\bf D}$ of the dipole moment operator, where ${\bf x}_k$ is a unit vector along the $k$th Cartesian axis of a spatial coordinate system. From (5.45) of OPA \cite{OPA} we find that for pumping through the first excited $^2$P$_J$ state, the squared amplitude of the dipole operator that occurs in the denominator of (\ref{rpp12}) is given in terms of the spontaneous, radiative decay lifetime $\tau^{\{e\}}$ of the excited atom by
\begin{equation}
|\mathcal{D}|^2 =\frac{\hbar c^3(2J+1)}{4\tau^{\{e\}}\omega^{\{eg\}3}}.
\label{rpp14}
\end{equation}

\emph{Net optical pumping.}
\quad
The sum of the depopulation pumping (\ref{dpp8a})  and repopulation pumping (\ref{rpp2}) is the net optical pumping,
\begin{equation}
\frac{\partial}{\partial t}(x_k|\chi_j)=-\gamma_{\rm mx}\sum_{l}\lvec s_j|A_{\rm op}(x_k)|s_l)(x_k|\chi_l).
\label{nop2}
\end{equation}
The amplitude of the net optical pumping operator $A_{\rm op}$ is
\begin{equation}
\tilde A_{\rm op}(x_k)=\sum_{\bar\mu\mu}\frac{\tilde R_{\rm op}^{\bar\mu\mu}}{\sigma x_k-z_{\bar\mu\mu}},\quad\hbox{with}\quad
\tilde R_{\rm op}^{\bar\mu\mu}=\tilde R_{\rm dp}^{\bar\mu\mu}-\tilde R_{\rm rp}^{\bar\mu\mu}.
\label{nop8}
\end{equation}
As shown in (6.72) of OPA \cite{OPA} the net optical pumping operator satisfies the constraint that no atoms are created or destroyed by optical pumping,
\begin{equation}
\lvec s_0|\tilde A_{\rm op}(x_r)=0 \quad\hbox{or}\quad \lvec s_0|\tilde R_{\rm op}^{\bar\mu\mu}=0.
\label{nop10}
\end{equation}
\section{Steady-State Solution}
Adding the contribution (\ref{rid16}) for relaxation in the dark to the contribution (\ref{nop2}) from net optical pumping we find
\begin{equation}
\frac{\partial}{\partial t}|\chi_j)=-K_j|\chi_j)
-\gamma_{\rm mx}\sum_{lr}|x_r)\lvec s_j|A_{\rm op}(x_r)|s_l)(x_r|\chi_l).
\label{ne2}
\end{equation}
For $j\ne 0$, the steady-state solution of (\ref{ne2}) is
\begin{equation}
|\chi_j)=-p_j\sum_{lk}G_j|x_k)\lvec s_j|A_{\rm op}(x_k)|s_l)(x_k|\chi_l).
\label{ne8}
\end{equation}
Here we have introduced the dimensionless Green's function
\begin{equation}
G_j=\gamma_{j\infty}K_j^{-1},
\label{ne10}
\end{equation}
and a dimensionless optical-pumping parameter for the $j$th spin mode,
\begin{equation}
p_j=\frac{\gamma_{\rm mx}}{\gamma_{j\infty}},
\label{ne12}
\end{equation}
proportional to the light intensity. The characteristic damping rate $\gamma_{j\infty}=\gamma_{\rm w}+\gamma_j+\gamma_{\rm vd}$ was given by (\ref{rid20}).
The parameter $p_j$ and the amplitude $|\chi_j)$ decrease as $|\gamma_{j\infty}|$ increases. For hyperfine coherences, the imaginary parts of $\gamma_{j\infty}$, the hyperfine Bohr frequencies, are so large compared to $|\gamma_{j\infty}|$ for non-hyperfine coherences, that it is a good approximation to neglect the hyperfine coherences entirely and retain only Zeeman coherences and population modes.
\subsection{Green's functions for multi-cusp collision kernels \label{mckernel}}
For the multicusp collision kernel of (\ref{ck14}) the damping rates (\ref{rid20}) become \cite{McGuyer12}
\begin{equation}
\gamma_{jn} = \gamma_{j0}+ \gamma_{\rm vd}\sum_k \frac{f_k n}{n + s_k}=\frac{N_j(n)}{D(n)}.
\label{gfc4}
\end{equation}
where $\gamma_{j0}=\gamma_{\rm w}+\gamma_j$. The denominator polynomial is simply
\begin{equation}
D(n) = (n+s_1)(n+s_2)\cdots(n+s_m).
\label{gfc6}
\end{equation}
Different spin modes may have different numerator polynomials
\begin{eqnarray}
N_j(n) &=& \gamma_{j0}D(n)+\gamma_{\rm vd}\sum_k\frac{f_k nD(n)}{n+s_k}\nonumber\\
 &=& \gamma_{j\infty} (n + r_{j1})(n+r_{j2})\cdots (n+r_{jm}),
\label{gfc8}
\end{eqnarray}
where the damping rate $\gamma_{j\infty}=\gamma_{\rm w}+\gamma_j+\gamma_{\rm vd}$ is the limit of (\ref{rid20}) for $n\to \infty$. Using (\ref{gfc4}) with (\ref{rid18}) and (\ref{ne10}) we find that the
multi-cusp Green's function is

\begin{eqnarray}
G_j &=& \sum_n\frac{\gamma_{j\infty}}{\gamma_{jn}}|v_n)\lvec v_n|=\gamma_{j\infty}\sum_n\frac{D(n)}{N_j(n)}|v_n)\lvec v_n|\nonumber\\
&=&1+ \frac{\gamma_{\rm vd}}{\gamma_{j0}}\overline{W}_j.
\label{gfc10}
\end{eqnarray}
The multi-cusp resolvent kernel is
\begin{equation}
\overline{W}_j=\sum_k g_{jk} C_{r_{jk}}.
\label{gfc12}
\end{equation}
Here $C_{r_{jk}}$ denotes a cusp kernel (\ref{ck8}) of sharpness $r_{jk}$, given as one of the roots of the numerator polynomial (\ref{gfc8}).
From a partial-fraction expansion of $D(n)/N_j(n)$ we find that the weights of the resolvent cusps are
\begin{equation}
g_{jk}=\lim_{n\to \,-r_{jk}}\frac{ \gamma_{j\infty} \gamma_{j0}(n+r_{jk}) D(n)}{\gamma_{\rm vd}r_{jk}N_j(n)}.
\label{gfc14}
\end{equation}
One can show that the fractional weights $g_{jk}$ sum to unity,
\begin{equation}
\sum_{k=1}^mg_{jk}=1.
\label{gfc16}
\end{equation}
An example of the pressure dependence of multicusp resolvent kernels is shown in Fig.~\ref{Fig2}. In the low-pressure limit, with $p\to 0$, the resolvent and collision kernels coincide, that is, $g_{jk}\to f_{k}$ and $r_{jk}\to s_k$. In the high-pressure limit with $p\to\infty$ all of the weight is transferred to the least sharp cusp, and the sharpness of this cusp approaches zero with increasing pressure,

\begin{equation}
r_{j1}\sim \frac{\gamma_{j0}}{\gamma_{j\infty}\sum_k(f_{k}/s_{k})}\propto\frac{1}{p}.
\label{gfc17}
\end{equation}
A cusp kernel with sharpness $r_{j1}\ll 1$ produces a Maxwellian distribution from any initial distribution of velocities.

\emph{Complex Green's functions.}\quad Since $\gamma_j^*=\gamma_{j^*}$, we see
the complex conjugate of the Green's function is
\begin{equation}
G_j^*=G_{j^*},
\label{gfc15}
\end{equation}
where $j^*$ is the conjugate index to $j$, defined by (\ref{hcm12}).
\subsection {Series solution}
For non-equilibrium modes with $j\ne 0$ we can write the mode amplitudes as a power series in the light-intensity parameter $p_j$ of (\ref{ne12}),
\begin{eqnarray}
|\chi_j)=\sum_{n=0}^{\infty}p_j^n|\chi^{\{n\}}_j).
\label{ss2}
\end{eqnarray}
The optical pumping parameter $p_j$ for the $j$th spin mode was given by (\ref{ne12}).
Special cases of (\ref{ss2}) that follow from (\ref{rid14}) are
\begin{eqnarray}
|\chi^{\{0\}}_j)=\delta_{j0}|v_0),\quad\hbox{and}\quad |\chi^{\{n\}}_0)=\delta_{n0}|v_0).
\label{ss3}
\end{eqnarray}
Substituting (\ref{ss2}) into (\ref{ne8}) we find for $j> 0$ and $n> 0$,
\begin{align}
|\chi_j^{\{n\}})
=&-\sum_{kl}G_j|x_k)\lvec s_j|A_{\rm op}(x_k)|s_l)\nonumber \\
&\quad\times(x_k|\chi_l^{\{n-1\}})\left(\frac{\gamma_{j\infty}}{\gamma_{l\infty}}\right)^{n-1}.
\label{ss4}
\end{align}

\emph{First-order spin polarization.}\quad
For the first-order spin mode, we can use (\ref{ss4}) with (\ref{ss3}) to find
\begin{equation}
|\chi_j^{\{1\}})=|\tilde \chi_j^{\{1\}})+|\tilde \chi_{j^*}^{\{1\}})^*,
\label{ss8}
\end{equation}
where
\begin{eqnarray}
|\tilde \chi_j^{\{1\}})&=&-\sum_{k}G_j|x_k)\lvec s_j|\tilde A_{\rm op}(x_k)|s_0)(x_k|v_0)\nonumber\\
&=&|\tilde \chi_{j{\rm w}}^{\{1\}})+|\tilde \chi_{j{\rm g}}^{\{1\}}).
\label{ss10}
\end{eqnarray}
We have used (\ref{gfc10}) to write the amplitude as the sum of a part coming from atoms that have had only wall collisions,
\begin{equation}
(x_k|\tilde \chi_{j{\rm w}}^{\{1\}}) = -\sum_{\bar\mu\mu}\frac{\lvec s_j|\tilde R_{\rm op}^{\bar\mu\mu}|s_0)(x_k|v_0)}{ x_k-z_{\bar\mu\mu}},
\label{ss12}
\end{equation}
and a background or ``pedestal" from atoms that have had gas-phase collisions,
\begin{equation}
(x_i|\tilde \chi_{j{\rm g}}^{\{1\}})=\frac{\gamma_{\rm vd}}{\gamma_{j0}}\sum_k (x_i|\overline{W}_{j}|x_k)(x_k|\tilde\chi_{j{\rm w}}^{\{1\}}).
\label{ss14}
\end{equation}
Physical insight can be gained by considering the first-order population shifts
\begin{equation}
|\chi_{\mu\mu}^{\{1\}})=\sum_j(\mu\mu|s_j)|\chi_j^{\{1\}}).
\label{ss16}
\end{equation}
Noting from (\ref{vev6}) that $\lvec v_0|\overline {W}_j=\lvec v_0|$, we multiply (\ref{ss14}) on the left by $\sum_i\lvec v_0|x_i)$ to find
\begin{equation}
\lvec v_0|\tilde \chi_{j{\rm g}}^{\{1\}})=\frac{\gamma_{\rm vd}}{\gamma_{j0}}\lvec v_0|\tilde\chi_{j{\rm w}}^{\{1\}}).
\label{ss18}
\end{equation}
According to (\ref{ss18}), the collisional background area $\lvec v_0|\tilde \chi_{j{\rm g}}^{\{1\}})$ and collision-free area $\lvec v_0|\tilde\chi_{j{\rm w}}^{\{1\}})$
are in the ratio of the velocity-damping rate due to gas-phase collisions, $\gamma_{\rm vd}$, to the gas-free spin damping rate due to wall collisions, $\gamma_{j0}=\gamma_{\rm w}+\gamma_j$.

\emph{Absorption rate.}\quad
Using (\ref{ss2}) we can write the absorption rate (\ref{dpp18}) as a power series in the light intensity,
\begin{equation}
\langle \delta{\Gamma}\rangle=\sum_{n}\langle \delta{\Gamma}^{\{n\}}\rangle,
\label{sr2}
\end{equation}
where
\begin{equation}
\langle \delta{\Gamma^{\{n\}}}\rangle=\langle \delta\tilde {\Gamma}^{\{n\}}\rangle+\langle \delta\tilde{\Gamma}^{\{n\}}\rangle^*,
\label{sr4}
\end{equation}
and the amplitude of the $n$th-order absorption rate is
\begin{equation}
\langle \delta\tilde{\Gamma}^{\{n\}}\rangle=\gamma_{\rm mx}\sum_{jk} p_j^n\lvec s_0|\tilde A_{\rm dp}(x_k)|s_j)(x_k|\chi_j^{\{n\}}).
\label{sr6}
\end{equation}
The zeroth-order rate is
\begin{equation}
\langle \delta\tilde{\Gamma}^{\{0\}}\rangle=\tilde\gamma_{\rm op},
\label{sr8}
\end{equation}
with the amplitude $\tilde\gamma_{\rm op}$ of the mean pumping rate given by (\ref{dpp22}).
The first-order rate is
\begin{equation}
\langle \delta\tilde{\Gamma}^{\{1\}}\rangle=\sum_{jk}\frac{\gamma_{\rm mx}^2}{\gamma_{j\infty}}\lvec s_0|\tilde A_{\rm dp}(x_k)|s_j)(x_k|\chi^{\{1\}}_j).
\label{sr10}
\end{equation}

For experiments where only population imbalances and no coherences are created by optical pumping and $\gamma_{\rm sd}$~$\ll$~$\gamma_{\rm vd}$, we can write (\ref{ss16}) as a sum of contributions from each ground-state energy sublevel $|\mu\rangle=|fm\rangle$,
\begin{equation}
\langle \delta\tilde{\Gamma}^{\{1\}}\rangle=\sum_{\mu k}\frac{\gamma_{\rm mx}^2}{\gamma_{0\infty}}\lvec s_0|\tilde A_{\rm dp}(x_k)|\mu\mu)(x_k|\chi^{\{1\}}_{\mu\mu}).
\label{sr12}
\end{equation}
For computing the absorption rate of a probe beam in a pump-probe experiment, $I_{\rm probe}$ should be used in (\ref{dpp2}) for computing $\gamma_{\rm mx}$, and the resulting $\langle \delta\tilde{\Gamma}^{\{1\}}\rangle$ must be multiplied by the ratio $I_{\rm pump}/I_{\rm probe}$.

Representative examples of the factors $\lvec s_0|\tilde A_{\rm dp}(x_k)|\mu\mu)$ and $(x_k|\chi^{\{1\}}_{\mu\mu})$ of (\ref{sr12}) are shown in Fig.~\ref{Fig4}.
\section{Comparison With Experiment\label{expt}}
The main purpose of this paper is to describe efficient ways to model the optical pumping of Na atoms under conditions similar to those of Na guidestar atoms. It has proven difficult to simulate conditions of Na guidestar atoms in laboratory experiments. It is relatively easy to carry out experiments at the same buffer-gas pressures as those experienced by guidestar atoms with non-reactive buffer gases like He, Ne, Ar, Kr, Xe and N$_2$. Unfortunately, the spin-damping rates of these gases are many orders of magnitude smaller than the velocity-damping rates. Because residual air at the $\sim$ 100~km altitude of the Na layer is still approximately 20\% O$_2$ molecules by volume and can even contain a few percent of dissociated O atoms, these paramagnetic species will cause the spin-damping and velocity-damping rates to be of comparable magnitude. Although Na atoms can have many binary collisions with O$_2$ molecules in the upper atmosphere, with negligible probability for a chemical reaction, the walls of a laboratory container quickly catalyze oxidation of Na atoms by O$_2$ gas. But the following examples show that the modeling methods work very well with laboratory VSOP experiments.

\subsection{Modulated Circular Dichroism of Na}
A particularly convenient way to investigate collisional effects on velocity-selective optical pumping was introduced by Aminoff {\it et al.}~\cite{Aminoff83}, who measured the effects of Ne buffer gas on the saturated absorption resonances of Na atoms. The Na in their experiments was held in a cylindrical glass cell and spin polarized by resonant, $589.8$~nm D1 laser light, pumping along a small, 1~G magnetic field with alternating circular polarization. In the limit of weak pumping light, this pumping scheme produces an alternating orientation (polarization of multipole index $L=1$). The polarization is detected as the modulated attenuation of a weak, counterpropagating probe beam of fixed circular polarization. We will call this a modulated circular dichroism (MCD) experiment.
\begin{figure}[ht]
\includegraphics[width=\columnwidth]{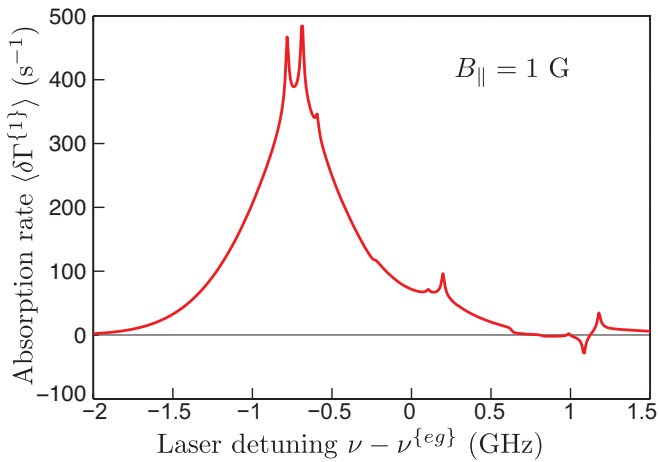}
\caption{(Color online) A MCD signal modeled with (\ref{sr12}) for measurements like those shown in Fig.~4 of Aminoff et al.~\cite{Aminoff83}. The modeling parameters were temperature $T=150~^\circ$C, optical-coherence linewidth $\gamma_{\rm oc}/2\pi = 10$~MHz, longitudinal field $B=1$~G, buffer-gas pressure $p=30$~mTorr, spatial diffusion coefficient $D_0=0.6$~cm$^2$\,s$^{-1}$, laser beam diameter $a=3.5$~mm, and a cell diameter $b=10$~mm. The cusp-kernel parameters of Fig. \ref{Fig2} were used with no adjustment.
\label{Fig5}}
\end{figure}
\begin{figure}[ht]
\includegraphics[width=\columnwidth]{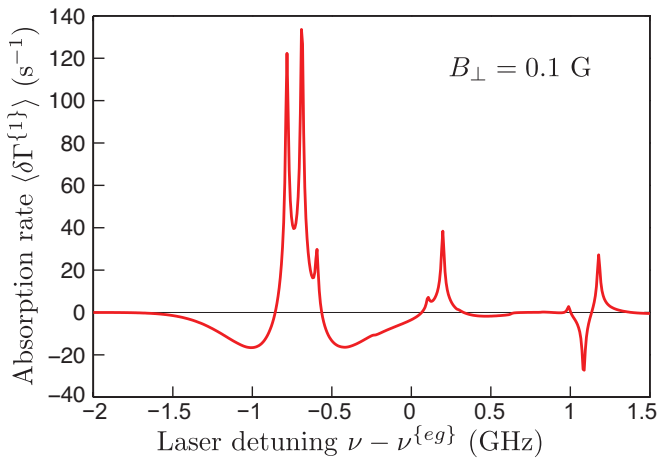}
\caption{(Color online) Magnetic depolarization signal predicted with (\ref{sr10}) for a transverse field of $B= 0.1$~G, but with all other modeling parameters the same as those of Fig.~\ref{Fig5}. The ``collisional background" of the saturated absorption signals is greatly modified and even reversed in sign. Such experiments could be used to determine the optimal cups kernels for describing velocity-changing collisions.
\label{Fig6}}
\end{figure}

In analyzing their MCD signals, Aminoff {\it et al.}~\cite{Aminoff83} found that KS kernels gave unsatisfactory fits to observations, and that much better fits could be obtained by adding a phenomenological, narrow Gaussian kernel, given by their Eq.~(32), to a broad KS kernel. The resulting two-term kernel does not produce a Maxwellian distribution of velocities in thermal equilibrium, that is, it does not satisfy the fundamental constraint (\ref{vev20}). The two-term kernel also cannot be inverted conveniently to obtain a Green's function, and in general it required the evaluation of a ``cumbersome double sum" \cite{Aminoff83}.

In Fig.~\ref{Fig5} we show the difference in absorption rates (\ref{sr10}) of the probe beam for right- and left-circularly-polarized pumping light. Of the 64 possible spin modes $|s_j)$, only two have amplitudes $|\chi_j)$ that differ appreciably from zero, the two independent modes with angular momentum $L=1$.  Computations are much easier for cusp or multicusp kernels than for the phenomenological kernel used by Aminoff {\it et al.}~\cite{Aminoff83}, because cusp kernels are easily inverted to give Green's functions. Even with no adjustment of the weights and sharpnesses of the three-cusp kernel, the modeled signal is very close to the observed signal shown in Fig.~4 of Aminoff {\it et al.}~\cite{Aminoff83} for 57 mTorr of Ne buffer gas. 

Fig.~\ref{Fig6} shows the modeled prediction of what would be observed
if the longitudinal field of Aminoff {\it et al.}~\cite{Aminoff83} were replaced by a transverse field of $0.1$ G and with other experimental conditions the same as those of Fig.~\ref{Fig5}. Quite different signals should be observed for magnetic fields large enough to cause substantial spin rotation before the optically pumped atoms can reach the cell walls. This is because a spin-polarized atom generated in one velocity group will rotate around the transverse magnetic field in the time needed to diffuse (in velocity space) to the velocity group detected by the circularly polarized probe beam. It is entirely possible for the spin polarization to rotate more than 90$^{\circ}$, reversing the sign of the probe signal in the process.
Magnetic fields with transverse components excite the 12 coherence modes with $\Delta m=\pm 1$, and if there are also longitudinal components of the magnetic fields the three population modes mentioned above are excited as well. So for non-longitudinal fields at least 15 modes $|s_j)$ and corresponding mode amplitudes $|\chi_j)$ need to included in the sums of (\ref{sr10}), and numerical calculations require more time. The nature of the ``magnetic
depolarization" signals generated by transverse magnetic fields is very sensitive to the parameters (sharpness, weight) used to construct the multicusp kernel. Systematic experiments with magnetic depolarization of the MCD signals at various buffer gas pressures and various magnetic fields, analyzed with the efficient modeling methods we have outline in this paper, would provide a very good way to determine the optimum parameters for cusp kernels.
\subsection{Velocity-selective optical pumping in potassium vapor}
\begin{figure}[t]
\includegraphics[width=\columnwidth]{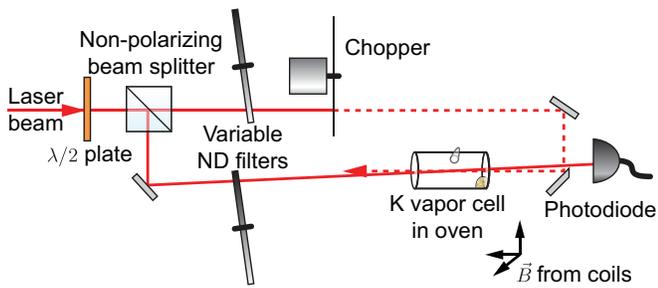}
\caption{(Color online) Schematic of a saturated-absorption experiment, described in more detail in the text. For the somewhat similar experiment of Aminoff {\it et al.} \cite{Aminoff83}, modeled in Fig.~\ref{Fig5}, and for the proposed experiment of Fig.~\ref{Fig6}, the intensity chopper is replaced by a device that modulates the circular polarization of the pumping light, and a weak probe beam has fixed circular polarization. Alternatively, one could
modulate the polarization of the probe beam and use fixed circular polarization for the pump beam, which would eliminate any concerns about transient buildup and decay of the optically pumped spin polarization.\label{Fig7}}
\end{figure}

In our own laboratory, we have completed preliminary experiments on velocity-selective optical pumping with K vapor. The basic experimental arrangement, depicted in Fig.~\ref{Fig7}, is almost identical to that of earlier experiments by Bloch {\it et al.}~\cite{Bloch96} which were done with no buffer gas. The pump beam was produced by a diode laser and was frequency-scanned across the 770 nm D1 absorption line of K atoms. The probe beam was a retro-reflected fraction of the pump beam. The pump-beam intensity was reduced with neutral density filters to keep it well within the first-order regime where (\ref{sr12}) describes the attenuation of the probe light. To ensure that optical pumping by the the probe beam was negligible, the probe intensity was attenuated by about about a factor of ten with respect to the pump beam. Both the pump and probe beams were linearly polarized along a 1~G magnetic field. The approximate radius of the beams was $a=1$ mm. The pump beam intensity was modulated on and off at 80~Hz with a chopper wheel. The pump beam produced a modulated spin polarization of the atoms which modulated the transmitted intensity of the probe beam. The intensity modulation of the transmitted probe beam was detected with a lock-in amplifier, referenced to the chopper wheel.

\begin{figure}[ht]
\includegraphics[width=8.3cm]{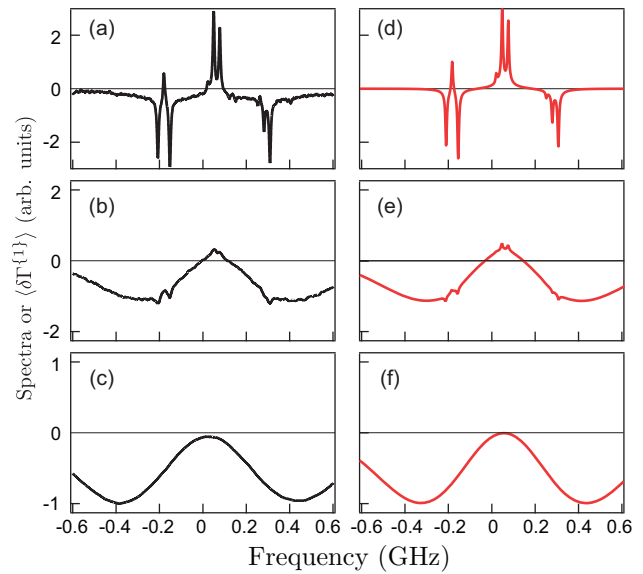}
\caption{(Color online) The left panels are saturated absorption signals from potassium vapor, measured with the apparatus of Fig.~\ref{Fig7}. The right panels are signals modeled with (\ref{sr12}) for monoisotopic $^{39}$K vapor. The buffer-gas pressures were: (a) No intentionally added buffer gas, (d) 1~mTorr of N$_2$; (b) or (e) 100~mTorr of Kr, (c) or (f) 1500~mTorr of N$_2$; The modeling parameters were: diffusion coefficient at 1~atm $D_0= 0.10$~cm$^2$\,s$^{-1}$ for both Kr and N$_2$, laser beam diameter $a=1$~mm, cell diameter $b=5$~mm. The cusp-kernel parameters of Fig. \ref{Fig2} were used with no adjustment.
 \label{Fig8}}
\end{figure}

Our glass cells included both spheres and cylinders, but for modeling we took a representative cylinder with a radius $b = 5$~mm. Before filling, a few drops of K metal were distilled under vacuum into the cell. The metal was of natural isotopic abundance, 93.26\% $^{39}$K and 6.73\% $^{41}$K, and the relatively small signals from $^{41}$K can be seen in some experimental data. Most cells were filled with low pressures of Kr or N$_2$ gas but some ``vacuum" cells had no intentional gas. Some of these vacuum cells were found to have small amounts of gas in them, as was shown clearly by collisional pedestals in the VSOP signals. During the experiments, the K vapor pressure was controlled by keeping the cell at a temperature near $50~^\circ$C in a temperature-stabilized oven with glass windows. The optical depth of the vapor was intentionally kept low (about 0.1 at the peak of the Doppler profile) in order to keep the laser intensity relatively uniform down the entire length of the cell. As shown in Fig.~\ref{Fig8}, the dominant $^{39}$K signals observed with the apparatus of Fig.~\ref{Fig7} could be simulated very well with the first-order absorption rate of (\ref{sr12}), with the parameters mentioned in the figure caption.
Because of the experimental arrangement, which cannot excite coherences, only the 8 population modes out of the 64 possible spin modes $| s_j )$ have amplitudes $| \chi_{\mu\mu} )$ that differ appreciably from zero.

\appendix*
\section{Spatial diffusion and velocity damping \label{sdvd}}
Here we show that the velocity damping rate $\gamma_{\rm vd}$ can be inferred from the measured spatial diffusion rate $D$ and the smallest, non-zero eigenvalue, $\alpha_1$ of the collision operator $A_{\rm vd}$ with the formula
\begin{equation}
\gamma_\text{vd} = \frac{v_D^2}{2 \alpha_1 D}.
\label{sd80}
\end{equation}
This is consistent with Eq.~(2.28) of Berman, Haverkort and Woerdman~\cite{Berman86}. 

Let $u$ be the spatial position in units of $v_D/\gamma_{\rm vd}$, the characteristic distance an atom can travel in one velocity-damping time, $1/\gamma_{\rm vd}$. The dimensionless time is $\tau=t\gamma_{\rm vd}$, where $1/\gamma_{\rm vd}$ is the characteristic time between velocity-changing collisions.
We ignore the atomic spin and consider atoms with a number density $N = N(u,x,\tau)= \sum_{n=0}^\infty N_{n}(u,\tau) (x|v_n)$, such that at the time $\tau$ the probability of finding a particle with position between $u$ and $u+du$ and velocity between $x$ and $x+dx$ is $N(u,x,\tau)du dx$. If we quantize the velocity $x$, we can write the density $N$ as an abstract column vector
\begin{equation}
|N) = \sum_n N_n|v_n),
\label{sd2}
\end{equation}
with velocity-mode amplitudes $N_n = \lvec v_n | N )$.
We assume that $|N)$ evolves according to the Boltzmann equation in one dimension,
\begin{equation}
\frac{\partial}{\partial \tau}|N)=-\left \{ \frac{\partial}{\partial u} X + A_{\rm vd} \right\}|N),
\label{sd4}
\end{equation}
The collision operator $A_{\rm vd}$ was given by (\ref{vev36}). We may use the identity $xH_n=nH_{n-1}+H_{n+1}/2$ and the definition (\ref{ks6})
of the velocity modes to write the velocity operator $X$ in (\ref{sd4}) as
\begin{equation}
X=\sum_{n=0}^\infty \sqrt{\frac{n+1}{2}}\bigg\{|v_{n+1})\lvec v_n|+|v_n)\lvec v_{n+1}|\bigg\}.
\label{sd6}
\end{equation}
Substituting (\ref{sd6}) and (\ref{sd2}) into (\ref{sd4}), multiplying the resulting equation on the left by $\lvec v_0|$ and $\lvec v_1|$, and remembering that
$\alpha_0=0$ we find
\begin{eqnarray} \frac{\partial N_0}{\partial \tau} &=& - \sqrt{\frac{1}{2}} \frac{\partial N_{1}}{\partial u} \label{sd8}
\\ \frac{\partial N_1}{\partial \tau} &=& - \sqrt{\frac{1}{2}} \frac{\partial N_{0}}{\partial u} - \frac{\partial N_2}{\partial u} - \alpha_1 N_1.
\label{sd10}
\end{eqnarray}
We are interested in ``late-time" distributions $N$ that have Maxwellian, or very nearly Maxwellian velocity distributions, where
the mode amplitudes $N_n = \lvec v_n | N )$ decrease rapidly with $n$ so that $N_0 \gg N_1 \gg N_2 \gg \cdots$. Retaining only the first two amplitudes,
$N_0$ and $N_1$, multiplying (\ref{sd8}) by $\partial/\partial \tau $, multiplying (\ref{sd10}) by $\partial /\partial u$, and combining the results
to eliminate $N_1$ we find
\begin{equation}
\frac{\partial^2 N_0}{\partial \tau^2} = \frac{1}{2} \frac{\partial^2 N_0}{\partial u^2} - \alpha_1 \frac{\partial N_0}{\partial \tau}.
\label{sd12}
\end{equation}
The distribution $N_0$ will evolve more and more slowly with increasing time, so for sufficiently late time we can ignore $\partial^2 N_0/\partial \tau^2$ compared to $\partial N_0/\partial \tau$ in (\ref{sd12}) and find the diffusion equation
 \begin{equation} \frac{\partial N_0}{\partial \tau} = \frac{1}{2 \alpha_1} \frac{\partial^2 N_0}{\partial u^2}, \end{equation} with the diffusion coefficient (in dimensionless units) \begin{equation} D=\frac{1}{2\alpha_1}.
\label{sd14}
\end{equation}
For a cusp kernel with sharpness $s$ we see from (\ref{ck7}) that $\alpha_1=1/(s+1)$, so (\ref{sd14}) gives $D=(s+1)/2$. For large sharpnesses, the diffusion coefficient is very nearly half the value of the sharpness, which is intuitively reasonable since sharp kernels represent velocity damping dominated by grazing-incidence collisions, which do little to hinder spatial diffusion. Using the unit of length $v_D/\gamma_{\rm vd}$ and unit of time $1/\gamma_{\rm vd}$ to convert the dimensionless diffusion coefficient (\ref{sd14}) to dimensional units (cm$^2$\,s$^{-1}$ for cgs units) we find (\ref{sd80}).

{\bf Acknowledgements.}
 The authors are grateful to M. J. Souza for making cells, and to Natalie Kostinski and Ivana Dimitrova for contributions to the experimental apparatus. This work was supported by the Air Force Office of Scientific Research.


\begin{thebibliography}{99}
\bibitem{Happer94} W. Happer, G. J. MacDonald, C. E. Max and F. J. Dyson, J. Opt. Soc. Am. {\bf 11}, 263 (1994).
\bibitem{Aminoff82} C. G. Aminoff and M. Pinard, J. Physique {\bf 43}, 263 (1982).
\bibitem{Quivers86} W. W. Quivers, Jr., Phys. Rev. A, {\bf 34}, 3822 (1986).
\bibitem{Tomasi93} F. de Tomasi, M. Allegrini, E. Arimondo, G.S. Agarwal and P. Ananthalakshmi, Phys. Rev. A {\bf 48}, 3820 (1993).
\bibitem{Bouchiat63a} M. A. Bouchiat, J. Physique {\bf 24}, 379 (1963).
\bibitem{Bouchiat63b} M. A. Bouchiat, J. Physique {\bf 24}, 611 (1963).
\bibitem{McGuyer12} B. H. McGuyer, R. Marsland III, B. A. Olsen and W. Happer, Phys. Rev. Lett. {\bf 108}, 183202 (2012).
\bibitem{Snider86} R. F. Snider, Phys. Rev. A {\bf 33}, 178 (1986).
\bibitem{Morgan10} S. W. Morgan and W. Happer, Phys. Rev. A {\bf 81}, 042703 (2010).
\bibitem{Keilson52} J. Keilson and J. E. Storer, Q. Appl. Math. {\bf 10}, 243 (1952).
\bibitem{Weideman94} J. A. C. Weideman, SIAM J. Numer. Anal. {\bf 34}, 1497 (1994).
\bibitem {OPA} W. Happer, Y-Y. Jau and T. G. Walker, {\it Optically Pumped Atoms}, Wiley-VCH GmbH Verlag, Weinheim (2010).
\bibitem{Ernst}R. R. Ernst, G. Bodenhausen and Alexander Wokaun, {\it Principles of Nuclear Magnetic Resonance in One
and Two Dimensions}, Oxford University Press (1990).
\bibitem{Gibble91} K. E. Gibble and A. Gallagher, Phys. Rev. A {\bf 43}, 1366 (1991).
\bibitem{Ho86} Tak-San Ho and Shih-I Chu, Phys. Rev. A {\bf 33}, 3067 (1986).
\bibitem{Wu88} Z. Wu, S. Schaefer, G. D. Cates and W. Happer, Phys. Rev. A {\bf 37}, 1161 (1988).
\bibitem{Abramowitz} M. Abramowitz and I. Stegun, {\it Handbook of Mathematical Functions}, Dover Publications, New York (1965).
\bibitem{Aminoff83} C. G. Aminoff, J. Javanainen and M. Kaivola, Phys. Rev. A {\bf 28}, 722 (1983).
\bibitem{Bloch96} D. Bloch, M. Ducloy, N. Senkov, V. Velichansky and V. Yudin, Laser Physics {\bf 6}, 670 (1996).
\bibitem{Berman86} P. R. Berman, J. E. M. Haverkort and J. P. Woerdman, Phys. Rev. A {\bf 34}, 4647 (1986).





\end{thebibliography}
\end{document}